\documentstyle[11pt,epsfig]{article}
\begin{document}
 \title{ Random Phase Approximation  and extensions applied to a bosonic field theory}
\author{ H. Hansen$^{(1)}$, G. Chanfray$^{(1)}$, D.  Davesne$^{(1)}$ and 
P. Schuck$^{(2)}$\\   
$^{(1)}${\small\it IPN Lyon,  43 Bd  du 11
Novembre  1918, F-69622 Villeurbanne Cedex} \\
$^{(2)}${\small\it IPN Orsay, Universit\'e Paris-Sud, France}} 
%\begin{center}
%\footnote{IPN Lyon, 43 Bd du 11 Novembre 1918, F-69622 Villeurbanne Cedex} 
%\end{center} 
\maketitle 
\begin{abstract} 
%\vspace*{50mm}
 
An application of a self-consistent version of RPA to quantum field theory 
with broken symmetry is presented. Although our approach can be applied to any 
bosonic field theory, we specifically 
study the $\varphi^4$ theory in 1+1 dimensions.
We show that standard RPA approach leads to an instability which can be removed 
when going to a superior version, {\it i.e.} the renormalized RPA. We present a method
based on the so-called charging formula of the many electron problem to calculate
the correlation energy and the RPA effective potential.
\end{abstract}

%%%%%%%%%%%%%%%%%%%%%%%%%%%%%%%%%%%%%%%%%%%%%%%%%%%%%%%%%%%%%%%%%%%
\section{Introduction}
%%%%%%%%%%%%%%%%%%%%%%%%%%%%%%%%%%%%%%%%%%%%%%%%%%%%%%%%%%%%%%%%%%%
One  central aim of the relativistic heavy ion program is to produce 
highly excited hot and dense matter possibly constituting  a quark-gluon 
plasma. In such a phase quarks and gluons should be liberated and chiral 
symmetry should be realized in its Wigner form. A central theoretical 
question is thus to have a correct description of the broken vacuum and of 
the progressive restoration of chiral symmetry with increasing temperature
and/or baryonic density. This problem is highly 
non-perturbative in nature  and usual perturbative loop expansion technique
are certainly not sufficient by construction. These features provide at least one
important motivation to develop tractable non perturbative methods to be applied in
the context of (effective) field theories with broken symmetries.  

\noindent
One hope is to try to adapt to quantum field 
problems  very well controlled non perturbative methods  from the nuclear 
many-body problem possibly of variational nature. A very popular method, 
known as the Gaussian approximation for interacting bosons, 
exactly corresponds to the Hartree-Fock-Bogoliubov mean field approximation
which constitutes 
the basic building block of the nuclear 
many-body theory {\it i.e.} for fermions, see e.g. \cite{RS80}. 
This variational method has already been
applied to  theories with a global symmetry \cite{S84} as well as with a local symmetry
\cite{KV89}. 
Nevertheless when this HFB approach is applied to a bosonic $O(N)$
model ({\it i.e.} to the linear sigma model) the pion appears with a finite
mass or, in other words, Goldstone theorem is violated as discussed in \cite{DMS96}. 
As demonstrated in
recent papers, RPA fluctuations are able to generate this soft mode, known as 
the spurious mode in the context of nuclear physics \cite{ACSW96,ASW97}. 
The RPA  approach thus appears
as a very promising technique to treat non-perturbative problems in the context of
quantum field theory.   
However in its simplest form this method is not
variational and its predictions are not always very well under control. In
particular, it is well-known that standard RPA has the tendency to overestimate
the attractive correlation energy, at least in examples of nuclear physics. 
The purpose of this paper is therefore two-fold: 
first, a complete presentation  of the RPA technique is given
in the context of the simplest field theory, namely the $\lambda \varphi^4$ theory
with a specific application to the  $1+1$ dimensional case. The second goal 
is to develop the formalism of more elaborated versions of the RPA approach, 
namely the so-called renormalized RPA (r-RPA), which, to our
knowledge, has never been done before. Let us mention 
that there exist previous attempts
to apply many-body techniques  for this specific 
problem. For instance, H\"{a}user {\it et al.}  
use the  cluster expansion of Green's functions \cite{HCPT95}.
Of course other methods such as lattice calculation have been employed, 
see {\it e.g.} \cite{JS94,LW97,MRL99}. 
Our aim is not really to compete with these numerical methods but 
actually to develop a tractable approach allowing direct physical interpretations 
in view of further applications in the
context of chiral or gauge theories. In this  preliminary work our aim is
rather modest and limited to the presentation of the formalism and the discussion 
of the remaining
problems to be solved such as the explicit covariance of the obtained results.
We also present the method to calculate the 
RPA correlation energy in the context of a field theory within a Green's function
formalism. One important point of this paper is to adapt the so called 
``charging formula''  method  (see e.g. \cite{FW}) 
for quantum field theory  beyond the standard RPA scheme.
Some  numerical results are also   obtained showing in particular how to
cure the instability of the  standard RPA in the broken symmetry region.

%%%%%%%%%%%%%%%%%%%%%%%%%%%%%%%%%%%%%%%%%%%%%%%%%%%%%%%%%%%%%%%%%%%
\section{The  $\varphi^4$ theory}

%%%%%%%%%%%%%%%%%%%%%%%%%%%%%%%%%%%%%%%%%%%%%%%%%%%%%%%%%%%%%%%%%%%

\subsection{The Hamiltonian}

We consider the Lagrangian density ~: 
\begin{equation}
{\cal L}={1\over 2}\,\partial^\mu\varphi(x)\, \partial_\mu\varphi(x)\,-
{1\over 2}\,\mu_0^2\,\varphi^2(x)\,-\,{b\over 24}\varphi^4(x)
\end{equation}
where $\mu_0^2$ is a constant and the bare coupling constant $b=\lambda/6$
is positive for  reasons of stability. We decompose the scalar field 
$\varphi(x)$ in a classical part
or condensate $s$ and a fluctuating piece $\phi(x)$~:
\begin{equation}
\varphi(x)=\phi(x)\,+\,s,\qquad s=\langle\varphi(x)\rangle~.
\end{equation}
The presence of the condensate $s$ indicates a spontaneous breaking  of the
underlying $\varphi\to -\varphi$ symmetry. Introducing the conjugate field
$\Pi(x)$, one obtains for the Hamiltonian (in d+1 dimensions)~: 
\begin{eqnarray}
H = & &\int d^d x\,\,\bigg\{{1\over 2} \mu^2_0\,s^2\,+\,{b\over 24}\,s^4\,+\, 
\left(\mu_0\,s\,+\,{b\over 6}\,s^3\right)\,\phi(x)\,\nonumber\\
 & & + \,{1\over 2}\big[\Pi^2(x)\,+\,\left(\partial_i\phi\right)^2(x) \,+\,\big(
\mu^2_0\,+{b\over 2}\,s^2\big)\,\phi^2(x)\big]\nonumber\\
& & +\,{b\,s\over 6}\,\phi^3(x)\,+\,{b\over 24}\,\phi^4(x)\bigg\}~.
\end{eqnarray}
Putting the system in a large box of volume $V=L^d$, 
it is convenient to work in momentum space and to expand the fields according to~:
\begin{equation}
\phi(x)={1\over \sqrt{V}}\,\sum_{\vec q}\,e^{i\vec q\cdot \vec x}\,\phi_{\vec q}(t)\quad,
\qquad
\Pi(x)=-\,{i\over \sqrt{V}}\,\sum_{\vec q}\,e^{i\vec q\cdot \vec x}\,\Pi_{\vec q}(t)~.
\end{equation}
The hermiticity of $\phi(x)$ and $\Pi(x)$ imposes $\phi^\dagger_{\vec q}
=\phi_{-\vec q}$ and $\Pi^\dagger_{\vec q}=-\Pi_{-\vec q}$ and canonical equal-time 
commutation relations translate into~: 
\begin{equation}
\big[\phi_{\vec q}\, ,\,\Pi^\dagger_{\vec q'} \big]=\delta_{\vec q, \vec q'}~.
\end{equation}
The Hamiltonian can be rewritten as~:
\begin{eqnarray} 
H = & & V\left({1\over 2} \mu^2_0\,s^2\,+\,{b\over 24}\,s^4\right)\,+\,
\sqrt{V}\,\sum_{\vec q}\,\left(\mu_0\,s\,+\,{b\over 6}\,s^3\right)\,
\phi_{\vec q}\,\delta_{{\vec q}, 0}\nonumber\\
& & +\sum_{\vec q}\,\left(\Pi_{\vec q}\,\Pi^\dagger_{\vec q}\,+\,
{\cal O}^2_{q}\,\phi_{\vec q}\,\phi^\dagger_{\vec q}\right)\,+\,
{1\over 6}\,\sum_{123}\,V_{1,2,3}\,\phi_1 \,\phi_2\,\phi_3\nonumber\\
& & +{1\over 24}\,\sum_{1234}\,V_{1,2,3,4}\,\phi_1 \,\phi_2\,\phi_3\,\phi_4
\label{HSTART}
\end{eqnarray}
where $\phi_i$ is a short-hand notation for $\phi_{\vec q_i}$ and the 
 three-body and four-body interactions are given by~:
\begin{equation}
V_{1,2,3}={b\,s\over\sqrt{V}}\,\delta_{{\vec q_1}+{\vec q_2}+{\vec q_3}},\qquad
V_{1,2,3,4}={b\over V}\,\delta_{{\vec q_1}+{\vec q_2}+{\vec q_3} +{\vec q_4}}~.
\label{H3H4}
\end{equation}
The bare single particle energy appearing in the Hamiltonian is~:
\begin{equation}
{\cal O}^2_{q}={\vec q}^2\,+\,\mu^2_0\,+\,{b\over 2} \,s^2~.
\end{equation}
In the following we will call $H_3$ and $H_4$ the three-body and four-body interacting
Hamiltonians.

\subsection{The Gaussian approximation}

\begin{figure}
  \begin{minipage}[b]{0.48\linewidth}
    \centering\epsfig{figure=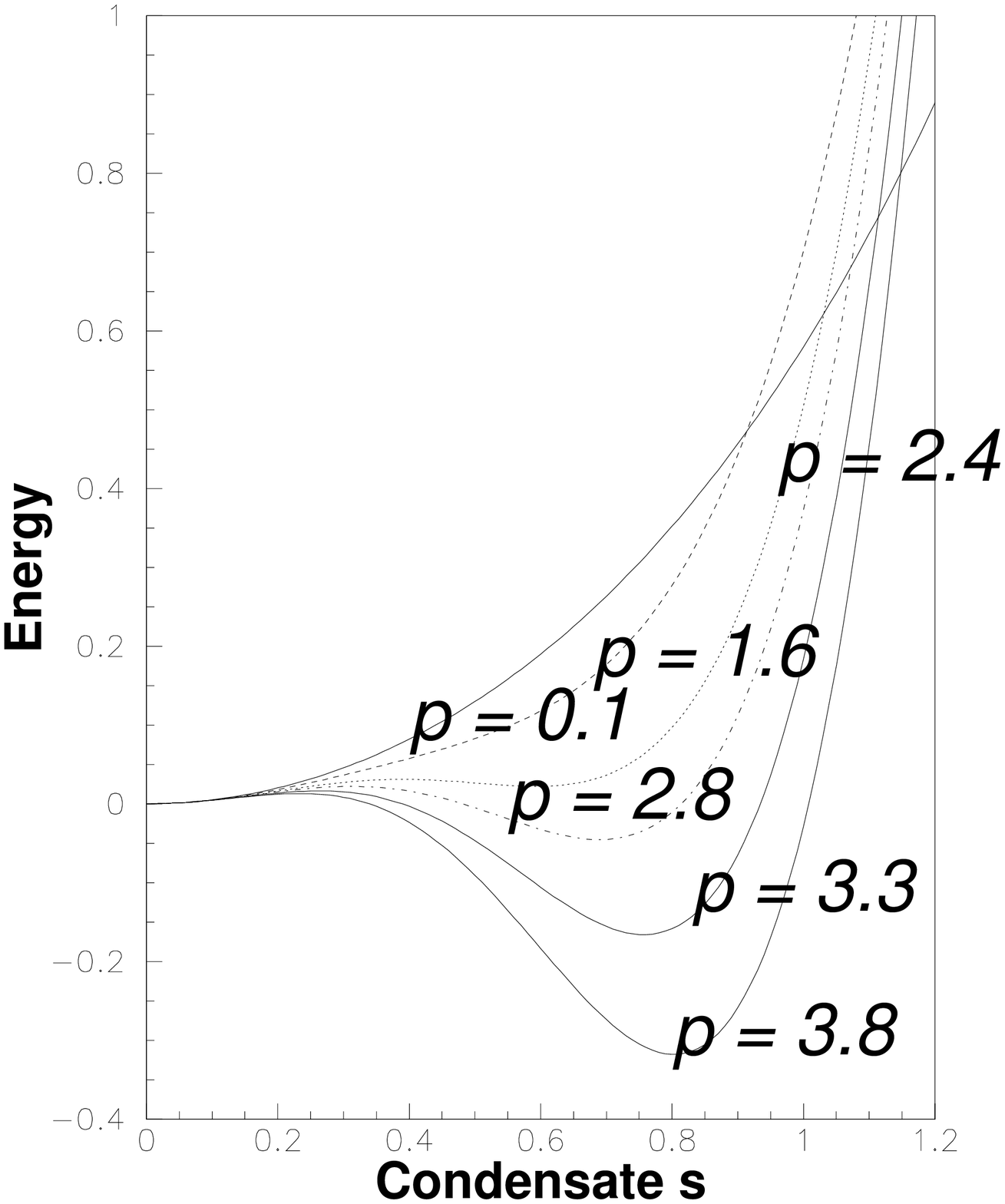,width=\linewidth}
  \end{minipage}\hfill
   \begin{minipage}[b]{0.48\linewidth}
    \centering\epsfig{figure=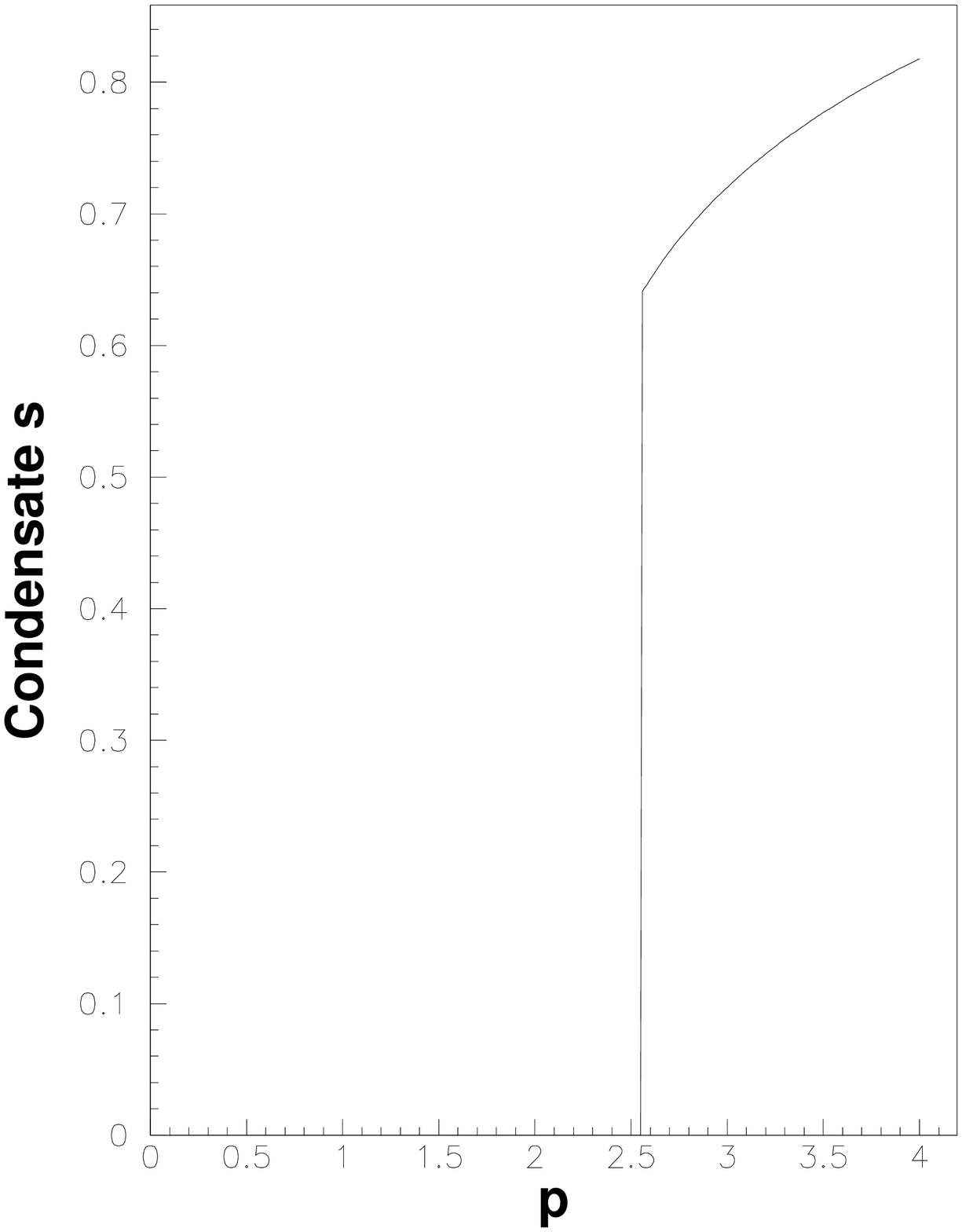,width=\linewidth}
  \end{minipage}
%\par\vspace*{-30mm}
\caption{
 Left panel~: Gaussian effective potential for various values of the
dimensionless coupling constant $p$. Right panel~: value of the condensate $s$ minimizing
the Gaussian effective potential as a function of $p$.}
\label{GAUSS}
\end{figure}

\begin{figure}
  \begin{minipage}[b]{0.48\linewidth}
    \centering\epsfig{figure=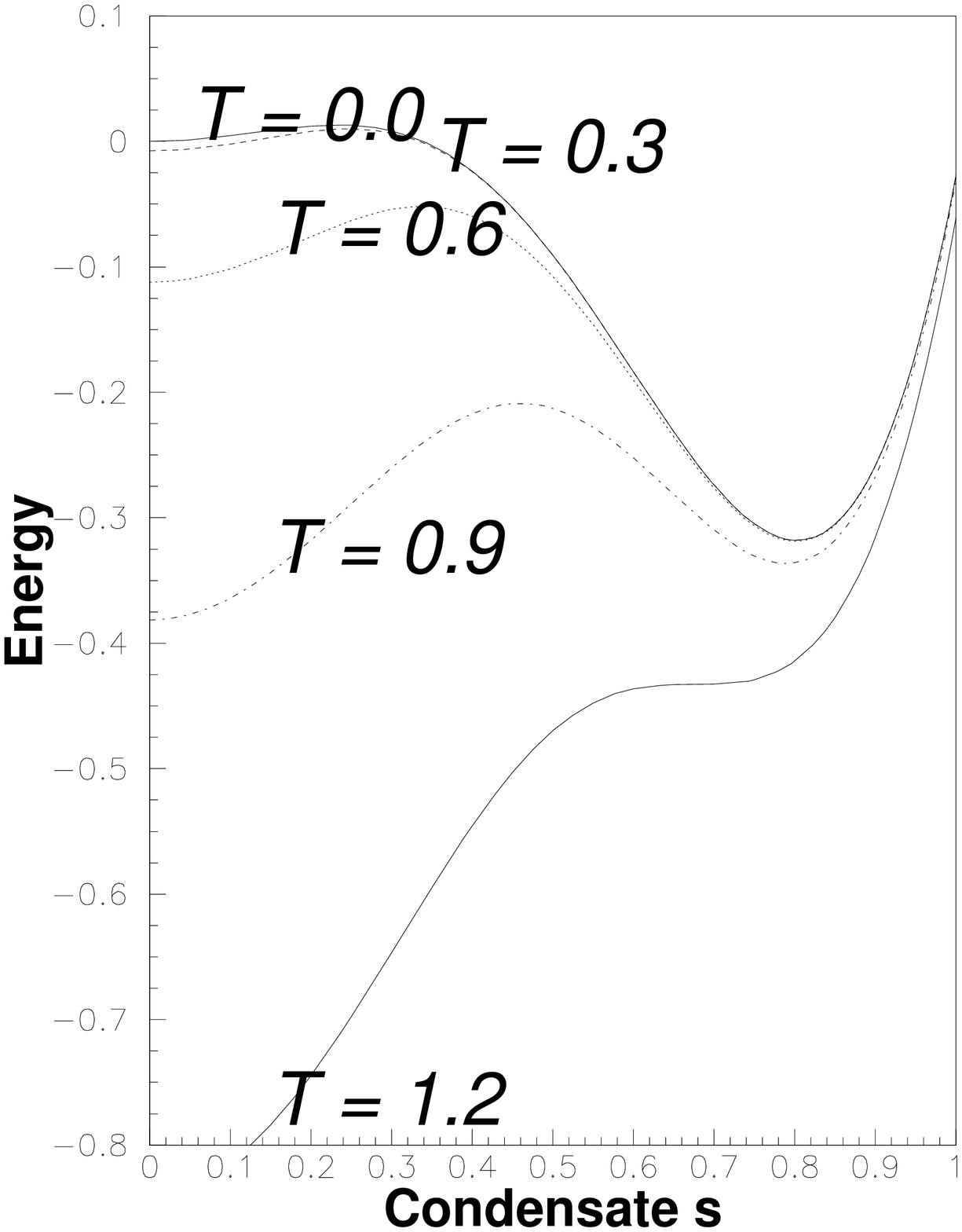,width=\linewidth}
  \end{minipage}\hfill
   \begin{minipage}[b]{0.48\linewidth}
    \centering\epsfig{figure=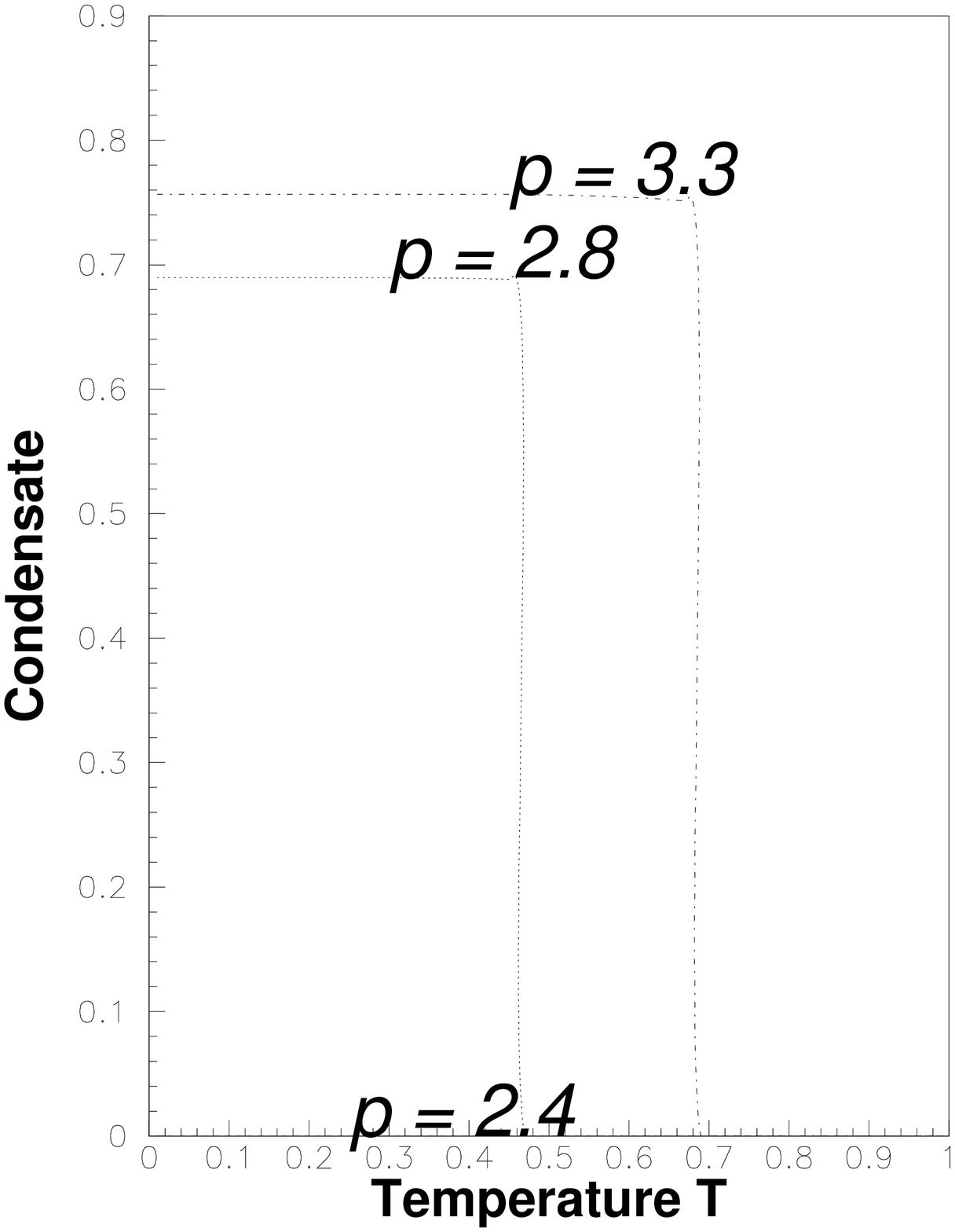,width=\linewidth}
  \end{minipage}
%\par\vspace*{-30mm}
\caption{
 Left panel~: Gaussian effective potential for $p=3.8$ for various
temperatures $T$ in units of $\mu$. Right panel~: value of the condensate $s$
minimizing
the Gaussian effective potential as a function of $T$ (in units of $\mu$) 
 for various values of $p$.}
\label{GAUSST}
\end{figure}

In the Gaussian approximation the ground state is represented by a trial  wave
function which is a functional of the field $\varphi({\vec x})$~:
\begin{equation}
|\psi(\varphi)\rangle={\cal N}\,\exp\left(-{1\over 2} \int d^dx\,d^d y\,
\big(\varphi({\vec x})-s\big)\,
h({\vec x}-{\vec y})\,\big(\varphi({\vec y})-s\big)\right)~.
\end{equation}
The optimal wave function is obtained by minimizing 
$\langle \psi(\varphi)| H |\psi(\varphi)\rangle$ with respect
to $h({\vec x}-{\vec y})$. The resulting function, which still depends on the
condensate $s$, defines the Gaussian effective potential. The various minima in $s$
correspond to the possible phases of the system. Working in momentum space, one
introduces the Fourier transform of  $h({\vec x}-{\vec y})$ according to~: 
\begin{equation}
h({\vec x}-{\vec y})=\int d^dq\, e^{i\,{\vec q}\cdot({\vec x}-{\vec y})}\,
\varepsilon_q~.
\end{equation}
It is easy to show that the trial ground state is the vacuum of the canonical 
destruction operator $b_{\vec q}$ such that~:
\begin{equation}
\phi_{\vec q}=\sqrt{1\over \ 2\varepsilon_q}\,\left(b_{\vec q}\,+\,
b^\dagger_{-\vec q}\right),\qquad
\Pi_{\vec q}=\sqrt{\varepsilon_q\over 2}\,\left(b_{\vec q}\,-\,
b^\dagger_{-\vec q}\right)~.
\end{equation}
The single particle excitation of this vacuum have  energies $\varepsilon_q$
which differ from the bare energies ${\cal O}_q$. In other words we have rotated the
original bare basis with single particle energies ${\cal O}_q$ into a basis 
associated to the $\varepsilon_q$'s through a Hartree-Fock-Bogoliubov (HFB)
transformation. In this HFB ground state the energy density is easily calculated by
using the Wick theorem~:
\begin{equation}
{\cal E}_0(\varepsilon, s)={1\over 2} \mu^2_0\,s^2\,+\,{b\over 24}\,s^4\,+\,
{1\over 2 V}\,\sum_{\vec q}\,\left({\varepsilon_q\over 2}\,+\,{{\cal O}^2_q\over
2\varepsilon_q}\right)\, +{b\over 8}\,\langle\phi^2\rangle^2\label{GAE1}
\end{equation}
where the scalar density $\langle\phi^2\rangle$ is given by~:  
\begin{equation}
\langle\phi^2\rangle={1\over V}\,\sum_{\vec q}\,\langle\phi_{\vec q}\,
\phi^\dagger_{\vec q}\rangle=
{1\over V}\,\sum_{\vec q}\,{1\over 2\varepsilon_q}
\equiv\int {d^d q\over (2\pi)^d}\,{1\over 2\varepsilon_q}~.
\end{equation}
Minimization with respect to $\varepsilon_q$ gives the HFB quasi-particle  mass $m$~:
\begin{equation}
m^2=\varepsilon_q^2-\,{\vec q}^2=\mu^2_0\,+{b\over 2}\,\big(s^2\,+\,
\langle\phi^2\rangle\big)~.
\end{equation}
It has been demonstrated in $d=3$ spatial dimensions that the above gap equation can
be rendered finite  by  appropriate mass and coupling constant renormalizations
\cite{KER}. 
Here we concentrate on to the $d=1$ case where the theory becomes
super-renormalizable and only requires a mass renormalization. We eliminate $\mu_0$ 
in favor of the renormalized mass $\mu$ according to~:
\begin{equation}
\mu^2_0=\mu^2\,-\,{b\over 2}\, \int_{-\Lambda}^{+\Lambda}\,{dq\over 2\pi}\,
{1\over 2\sqrt{q^2+\mu^2}}
\end{equation}
where $\Lambda$ is a ultraviolet cutoff. The gap equation becomes~~: 
\begin{equation}
m^2=\mu^2\,+\,{b\over 2}\,\big(s^2\,+\,
\langle\Delta\phi^2\rangle_\mu\big)\label{GAGA}
\end{equation}
with  
\begin{equation}
\langle\Delta\phi^2\rangle_\mu=\int_{-\Lambda}^{+\Lambda}\,{dq\over 2\pi}\,
\left({1\over 2\sqrt{q^2+m^2}}\,-\,{1\over 2\sqrt{q^2+\mu^2}}\right)= 
-{1\over 4\,\pi}\, \ln{m^2\over \mu^2}
\end{equation}
which is independent of the cutoff $\Lambda$. Re-injecting the solution for the mass
$m$ into the expression of the energy density one gets the effective potential
which is also finite, once the energy of the perturbative vacuum of
particles with mass $\mu$ is removed. The result is~:
\begin{equation}
{{\cal E}_0(s)\over \mu^2}={1\over 2}\,s^2\,+\,p\,s^4\,+\,
{1\over 8\pi}\,\left({m^2\over \mu^2}\,-1\,-{m^2\over \mu^2}\,
\ln{m^2\over \mu^2}\right)\,-\,{3\,p\over 16\pi^2}\,
\left(\ln{m^2\over \mu^2}\right)^2\label{GAE2} 
\end{equation}
where 
$$p=b/24\mu^2$$ 
is the dimensionless coupling constant. At low $p$ the effective
potential has a minimum at $s=0$~: the unbroken phase is thus stable. 
Increasing $p$, a new minimum develops at finite $s$
corresponding to a meta-stable broken phase. At a certain critical $p_c=2.57$
the deformed phase becomes stable and the system undergoes a first order phase
transition (see figure \ref{GAUSS}) in agreement with \cite{HCPT95}. The approach can be 
straightforwardly extended to finite temperature using the so-called statistical variational
principle. 
One obtains for the finite temperature effective potential 
({\it i.e.} the grand potential)~:
\begin{eqnarray}
{\Omega_0(s,T)\over \mu^2}&=& {T\over \mu}\,
\int{dq\over 2\pi}\,\ln\left(1-exp\left({\varepsilon_q \over T}\right)\right)\nonumber\\
& & +{1\over 2}\,s^2\,+\,p\,s^4\,+\,
{1\over 8\pi}\,\left({m^2\over \mu^2}\,-1\,-{m^2\over \mu^2}\,
\ln{m^2\over \mu^2}\right)
\, -\,3\,p\,\langle\Delta\phi^2\rangle_\mu^2(T)
\end{eqnarray}
with
\begin{equation}
\langle\Delta\phi^2\rangle_\mu(T)=\int\,{dq\over 2\pi}\,
\left({1\,+\,n(\varepsilon_q/ T)\over 2\sqrt{q^2+m^2}}\,
-\,{1\over 2\sqrt{q^2+\mu^2}}\right)
\end{equation}
where $n(x)=1/(exp(x)-1)$ is the Bose-Enstein distribution and the gap
equation (\ref{GAGA}) is unchanged once   $\langle\Delta\phi^2\rangle_\mu$ is replaced by its
finite temperature expression given just above.
Some numerical  results are shown on figure  \ref{GAUSST}. 
Choosing $\,p>p_c$ the vacuum is in a broken phase. At a certain critical temperature
$T_C$ one gets a first order  transition towards the symmetry restored phase. 

%%%%%%%%%%%%%%%%%%%%%%%%%%%%%%%%%%%%%%%%%%%%%%%%%%%%%%%%%%%%%%%%%%%%%%%%%%%%%%%%%%%%%%%%%
\section{The RPA approach}
%%%%%%%%%%%%%%%%%%%%%%%%%%%%%%%%%%%%%%%%%%%%%%%%%%%%%%%%%%%%%%%%%%%%%%%%%%%%%%%%%%%%%

\subsection{The equation of motion method (EOM)}

The aim of the RPA method is to describe the excitation spectrum of a Hamiltonian
$H$. The excited states $|\nu\rangle$ and the ground state $|0\rangle$ are defined by
the conditions~:
 \begin{equation}
|\nu\rangle= Q^\dagger_\nu\,|0\rangle,\qquad Q_\nu|0\rangle=0 ~.
\end{equation}
Minimizing $E_\nu=\langle\nu| H|\nu \rangle/\langle\nu|\nu \rangle$ with respect to the
operators $Q_\nu$, one gets the following set of equations \cite{RS80,R68}~: 
\begin{equation}
\langle 0|\bigg[\delta Q_\nu , \big[H, Q^\dagger_\nu\big]\bigg]|0\rangle=
\Omega_\nu\,\langle 0|\big[\delta Q_\nu , \ Q^\dagger_\nu\big]|0\rangle
\label{RPAEQ}\end{equation}
where $\Omega_\nu=E_\nu-E_0$ is the excitation energy. Equation (\ref{RPAEQ}) is 
supplemented by the conditions~:
\begin{equation}
\langle 0|\big[H, Q_\nu\big]|0\rangle=0
\label{COND}\end{equation}
which will determine the mean-field basis.  Eq.(\ref{COND}) is a natural complement 
to eq.(\ref{RPAEQ}), since both equations hold in the exact case \cite{DS90}. 
In general in this EOM approach, 
the excitation operators are searched only within a limited domain. They are 
 usually taken in the form~:
\begin{equation}
Q^\dagger_\nu=\sum_a\,\left(X^\nu_a\,A^\dagger_a\,-\,Y^\nu_a\,A_{-a}\right)  
\end{equation}
where the $A_a$ constitute a set of non-hermitian operators labeled by a set
 of quantum numbers $a$ (for instance momentum or isospin state).  
The RPA equations (\ref{RPAEQ}) 
now become matrix equations  which allow to determine the
excitation energy  and the $X$ and $Y$ amplitudes (summation over repeated indices is
understood)~: 
\begin{equation}
\pmatrix{{\cal A}_{a b}& {\cal B}_{a b}\cr
{\cal B}^*_{-a -b}&{\cal A}^*_{-a -b}}\,\pmatrix{X^\nu_b \cr Y^\nu_b}=
\Omega_\nu \, {\cal N}_{ab}\,\pmatrix{X^\nu_b \cr Y^\nu_b}
\end{equation}
with the matrix elements given by the double commutators~:
\begin{equation}
{\cal A}_{a b}=\langle 0|\bigg[A_a,\big[H, A^\dagger_b\big]\bigg]|0\rangle, \qquad
{\cal B}_{a b}=-\langle 0 |\bigg[A_a,\big[H, A_{-b}\big]\bigg]|0\rangle 
\end{equation}
and the norm matrix is~:
\begin{equation}
{\cal N}_{ab}=\pmatrix{\langle 0|\big[A_a, A^\dagger_b\big]|0\rangle &
\langle 0|\big[A_a, A_{-b}\big]|0\rangle\cr
-\langle 0|\big[A_a, A_{-b}\big]|0\rangle &
-\langle 0|\big[A_a, A^\dagger_b\big]|0\rangle}~.
\end{equation}
The previous conditions (\ref{COND}) 
translate into equations having the form of  gap equations:
\begin{equation}
\langle 0|\big[ H, A_a\big]|0\rangle=0\label{GAPEQ}~.
\end{equation}
In practice the quality of the RPA scheme depends on the approximation which are made.
Firstly the larger is the space of $A_a$ operators, the better is the
accuracy of the method. Here, in practice we will limit ourselves to single  and
pair boson operators namely $A^\dagger_a=\{ b^\dagger_{\vec q},\, 
b^\dagger_{\vec q}b^\dagger_{\vec q'}, \,  b^\dagger_{\vec q} b_{\vec q'}\}$. 
Secondly, one crucial point is the calculation of the ground state matrix elements 
of the double
commutators and of the matrix elements appearing in the gap equations. 
Calculating the matrix elements on the true
RPA ground-state constitutes the self-consistent RPA (SCRPA). In practice this is a
formidable task both formally and numerically, which only has been fully achieved in 
a very limited number of simple model cases. In the most common approximation 
({\it i.e.} the standard RPA), the ground state matrix elements entering the RPA
equations are calculated on the mean-field ground-state. 
In our case it coincides
with the Gaussian or HFB ground state discussed  above. There is an
intermediate approximation scheme, called in nuclear physics renormalized RPA 
(r-RPA)\cite{R68,H64,CDS94,TS95} 
which allows to incorporate part of the correlations in the ground state. 
In section 4, we will propose the first application of this method in the context
of quantum field theory. 

\subsection{The Dyson equation approach (DEA)}   

We present an alternative but equivalent formulation of  the  RPA scheme based on a 
Green's function (GF) approach \cite{SE73}. Although 
equivalent in principle to the equation of motion approach, this DEA method is more convenient
for quantum field theory. We define the time-ordered Green's functions~:
\begin{eqnarray}
G_{a,\bar b}(t,t')&=&-i\langle 0|T\left(A_a(t)\, ,\,A^\dagger_b(t')\right)|0\rangle
\nonumber\\
G_{a, -b}(t,t')&=&-i\langle 0|T\left(A_a(t)\, ,\,A_{-b}(t')\right)|0\rangle
\nonumber\\
G_{-\bar a, \bar b}(t,t')&=&-i\langle 0|T\left(A^\dagger _{-a}(t)\, ,
\,A^\dagger _{ b}(t')\right)|0\rangle\nonumber\\
G_{-\bar a, -b}(t,t')&=&-i\langle 0|T\left(A^\dagger _{-a}(t)\, ,
\,A_{-b}(t')\right)|0\rangle~.
\label{DEA29}
\end{eqnarray}
We introduce the energy representation of these GF according to~: 
\begin{equation}
G(t,t')=\int\,{dE\over 2\pi}\,e^{-i\,E\,(t-t')}\,G(E)
\end{equation}
and the matrix~:
\begin{equation}
{\cal G}_{a,b}(E)=\pmatrix{ G_{a,\bar b}(E) &G_{a, -b}(E)\cr
G_{-\bar a, \bar b}(E) & G_{-\bar a, -b}(E)}~.
\end{equation}
The RPA equations have now the form of a set of coupled integral equations 
for the various GF~: 
\begin{equation}
E\,{\cal G}_{a,b}(E)={\cal N}_{a,b}\,+\,
\sum_{c,d}\,\pmatrix{{\cal A}_{a c}& {\cal B}_{a c}\cr
{\cal B}^*_{-a -c}&{\cal A}^*_{-a -c}}\,{\cal N}^{-1}_{c,d}\,{\cal G}_{d,b}(E)
\label{RPAEQC}
\end{equation}
with~: 
\begin{equation}
{\cal A}_{a b}=\langle 0|\bigg[\big[A_a, H\big], A^\dagger_b\bigg]|0\rangle, \qquad
{\cal B}_{a b}=-\langle 0 |\bigg[\big[A_a,H\big], A_{-b}\bigg]|0\rangle~. 
\end{equation}
Notice that the double commutators are ordered differently from the ones obtained in the
equation of motion method. In practice, it can been shown that they are always identical. 

%%%%%%%%%%%%%%%%%%%%%%%%%%%%%%%%%%%%%%%%%%%%%%%%%%%%%%%%%%%%%%%%%%%%%%%%%%%%%%%
\section{The RPA and the renormalized RPA applied to the $\varphi^4$ theory}
%%%%%%%%%%%%%%%%%%%%%%%%%%%%%%%%%%%%%%%%%%%%%%%%%%%%%%%%%%%%%%%%%%%%%%%%%%%%%%%

\subsection{Solution of the RPA problem}

We will limit ourselves to the case where the $A$ operators are only one-body and two-body
operators. We introduce creation $b^\dagger_\beta$  and destruction 
operators $b_\beta$ depending on the parameters $\kappa_\beta$  according to~:
\begin{equation}
b_\beta=\sqrt{\kappa_\beta\over 2}\,\phi_\beta\,+\,
\sqrt{1\over 2 \kappa_\beta} \Pi_\beta
\end{equation}
or equivalently~:
\begin{equation}
\phi_{\beta}=\sqrt{1\over 2 \kappa_\beta}\,\left(b_\beta\,+\,
b^\dagger_{-\beta}\right),\qquad
\Pi_{\beta}=\sqrt{\kappa_\beta\over 2}\,\left(b_\beta\,-\,
b^\dagger_{-\beta}\right)~.\label{FIPI}
\end{equation}
By construction they obey standard canonical commutation relation for boson creation
and destruction operators~:
\begin{equation}
 [b_\beta,\, b^\dagger_{\beta'}]=\delta_{\beta,\beta'},\qquad
 [b_\beta,\,b_{\beta'}]=0~.
\end{equation} 
The $\beta$'s represent the quantum numbers of the created boson state; 
here this is simply a
momentum index and $-\beta$ represents the opposite momentum.

\smallskip\noindent 
The excitation operators $A^\dagger_a$ will be the one-body operators 
$b^\dagger_\alpha$ and the two-body
operators $b^\dagger_{\beta}\,b^\dagger_{\beta'}$ and $b^\dagger_{\beta}\,b_{-\beta'}$ 
 with $\beta\ne\beta'$ for the ``particle-hole'' operators. The gap equation
$\langle[H,b_\beta]\rangle=0$ will give the extrema in the condensate $s$ of the vacuum
energy. Since we will calculate  the effective potential giving directly the true
minimum ({\it i.e.} the stable phase) we will not use it here. The most interesting non
trivial gap equation is $\langle[H,b_\beta \, b_{-\beta}]\rangle=0$ which constrains the
basis, that is the $\kappa_\beta$ parameters. This gap equation gives 
($\langle ... \rangle$ stands for ground-state expectation value)~:
\begin{eqnarray}
\langle\Pi_\beta\Pi^\dagger_\beta\rangle\,-\,\varepsilon^2_\beta\,
\langle\phi_\beta\phi^\dagger_\beta\rangle &=&
{1\over 2} \sum_{1,2}\,\langle\phi_\beta \phi^\dagger_1\phi^\dagger_2
\rangle_{conn}\,V_{1,2,-\beta}\nonumber\\
& & +\,{1\over 6} \sum_{1,2,3}\,\langle\phi_\beta \phi^\dagger_1\phi^\dagger_2
\phi^\dagger_3\rangle_{conn}\,\label{GAPSC}
V_{1,2,3,-\beta}
\end{eqnarray}
where the suffix $conn$ means connected operators  which contain exclusively correlated
expectation values~:
\begin{equation}
\langle ABCD\rangle_{conn}=\langle ABCD\rangle\,-\,
\langle AB\rangle \langle CD\rangle\,-\,\langle AC\rangle \langle BD\rangle\,-\,
\langle AD\rangle \langle BC\rangle
\end{equation}
and $\varepsilon_\beta$ is given by~:
\begin{equation}
\varepsilon_\beta^2={\vec q}_\beta^2\,+\mu_0^2\,+\,{b\over 2} s^2\,+
{b\over 2}\langle\phi^2\rangle\label{GGAP}
\end{equation}
with~: 
\begin{equation}
\langle\phi^2\rangle={1\over V}\,\sum_{\vec q}\,\langle\phi_{\vec q}\,
\phi^\dagger_{\vec q}\rangle=
{1\over V}\,\sum_{\vec q}\,{1\over 2\kappa_q}
\langle 1\,+\,2 b^\dagger_{\vec q} b_{\vec q}\, +\,b^\dagger_{\vec q} 
b^\dagger_{-\vec q}\,+\,b_{\vec q} b_{-\vec q}\rangle\label{SCAL}~.
\end{equation}
The $\varepsilon_\beta$'s can be seen as the generalized mean field single-particle energies. But at
variance with the Gaussian case these energies  depend on the correlated Self-Consistent
scalar-density $\langle\phi^2\rangle$.

\medskip\noindent
{\it Standard RPA}

\smallskip\noindent
In the standard RPA, we omit the connected parts. We thus get from the gap equation
(\ref{GAPSC})~:
\medskip\noindent
\begin{equation}
\langle\Pi_\beta\Pi^\dagger_\beta\rangle\,-\,\varepsilon^2_\beta\,
\langle\phi_\beta\phi^\dagger_\beta\rangle=0\label{GAP0}
\end{equation}
and take for $\langle\phi_\beta\phi^\dagger_\beta\rangle$ its value in the Gaussian HFB
ground state. This implies (see eq.(\ref{FIPI})~:
\begin{equation}
\kappa_\beta=\varepsilon_\beta~.
\end{equation}
The basis is fixed and coincides with the HFB basis.  
Similarly all the matrix elements appearing in the calculation of the double
commutators and the norm matrix are simply obtained by using Wick theorem on the Gaussian
ground state.

\medskip\noindent
{\it Renormalized RPA}

\smallskip\noindent
In renormalized RPA (r-RPA), one still systematically omits 
correlated expectation values. Hence the
gap equation (\ref{GAP0}) remains valid. However, the scalar density   is not yet fixed. 
It has to be determined self-consistently. The problem of its evaluation (in a non
relativistic many-body problem for Fermi systems it corresponds to the occupation numbers)
is one of the  subtleties of self-consistent RPA (SCRPA). 
We will come to this problem later
on.  The gap equation can be rewritten as~:
\begin{equation} 
(\kappa^2_\beta\,-\,\varepsilon^2_\beta)\,\langle 1\,+\,2\,b^\dagger_\beta
\,b_\beta\rangle
=(\kappa^2_\beta\,+\,\varepsilon^2_\beta)\,\langle 
b^\dagger_\beta\,b^\dagger_{-\beta}\,+\,b_\beta\,b_{-\beta}\rangle~.
\end{equation}
The basis $\kappa_\beta$ is not yet totally fixed. However, 
we have checked that the result of the
r-RPA calculation does not depend on its choice, provided the above gap 
equation is satisfied. Thus we can choose it as
$\kappa_\beta=\varepsilon_\beta$ {\it i.e.} the generalized mean-field basis. This choice has
the merit of significantly simplifying the lengthy calculation of the double commutators.
In that case one has~:
\begin{equation}
\langle b^\dagger_\beta\,b^\dagger_{-\beta}\rangle=
\langle b_\beta\,b_{-\beta}\rangle =0~.
\end{equation}
In SCRPA this last property has to be always satisfied because the
$b^\dagger_\beta\,b^\dagger_{-\beta}$ operators are just linear combinations of the genuine
RPA excitation operators 
$Q^\dagger_\nu$ whose expectation values  on the true RPA ground state vanish by construction.

\noindent
The many-body operator expectation values are also calculated 
using Wick theorem but the resulting two-body operator matrix elements depend on the
various occupation numbers. In other words all the ground-state matrix elements 
entering the double-commutators and norm matrices are expressible in term
of the  fixed momentum densities   
${\cal N}_\beta=\langle\phi_\beta\phi^\dagger_\beta\rangle$. We display here some
examples~:
\begin{eqnarray}
\langle b_\beta b_{-\beta}\rangle&=&{\kappa^2_\beta-\varepsilon_\beta^2
\over 2\varepsilon_\beta}{\cal N}_\beta\nonumber\\
 \langle b_\beta^\dagger b_\beta \rangle&=& -{1\over 2}\,+\,
 {\kappa^2_\beta+\varepsilon_\beta^2\over 2\varepsilon_\beta}{\cal N}_\beta\nonumber\\
\langle b^\dagger_1 b^\dagger_2 b_3 b_4\rangle&=&
\langle b^\dagger_1 b_1\rangle \langle b^\dagger_2 b_2\rangle \,
(\delta_{1,3}\, \delta_{2,4}+\delta_{1,4}\,\delta_{2,3})\nonumber\\
 & &\,+
\langle b^\dagger_1 b^\dagger_{-1}\rangle \langle b_3 b_{-3}\rangle 
\,(\delta_{1,-2}\, \delta_{3,-4})~.
\end{eqnarray}

\medskip\noindent
We do not give  the details of the calculation to obtain the solution of the RPA
problem (\ref{RPAEQC}). Even if we limit ourselves to one- an two-body operators, it
represents quite a lot of algebra. The results for the various GF are
listed in the appendix. Here we give only the GF relative to the field
operator $\phi_{\vec P}$. For the one-particle GF {\it i.e.} the $\phi$ particle
propagator, one obtains~:
\begin{eqnarray}
G_{\phi_{\vec P}, \phi_{{\vec P}'}}(E) &\equiv & \delta_{{\vec P}, {\vec P}'}\,
G(E, {\vec P})\nonumber\\
G(E, {\vec P}) &= &\left(\,E^2\,-\,\varepsilon^2_{\vec P}\,-\, \Sigma(E,{\vec P})
\right)^{-1}\label{RESI}
\end{eqnarray}
where the mass operator has the following form~:
\begin{equation}
\Sigma(E,{\vec P})={b^2\, s^2\over 2}\,\tilde I(E,{\vec P})\, \qquad
\hbox{with} \qquad \tilde I(E,{\vec P})={I(E,{\vec P})\over 1\,-\,{b\over 2}
I(E,{\vec P})}\label{SEP}
\end{equation}
and the two-particle loop has the explicit expression~:
\begin{eqnarray}
I(E,{\vec P}) 
& = &\int \,{d{\vec k}_1\,d{\vec k}_2\over (2\pi)^d}\,\delta^{(d)}\left({\vec P}-
{\vec k}_1 - {\vec k}_2\right)\,\bigg[
{\varepsilon_1\,+\,\varepsilon_2\over 2\,\varepsilon_1\,\varepsilon_2}\,
{\varepsilon_1\,{\cal N}_1\,+\,\varepsilon_2\,{\cal N}_2\over 
E^2\,-\,\left(\varepsilon_1\,+\,\varepsilon_2\right)^2+i \eta}\nonumber\\
& &-\,
{\varepsilon_1\,-\,\varepsilon_2\over 2\,\varepsilon_1\,\varepsilon_2}\,
{\varepsilon_1\,{\cal N}_1\,-\,\varepsilon_2\,{\cal N}_2\over 
E^2\,-\,\left(\varepsilon_1\,-\,\varepsilon_2\right)^2+i \eta}\bigg]~.\label{IEP}
\end{eqnarray}
Notice that  correlations are present in this expression through the  densities 
${\cal N}_i=\langle\phi_i\phi^\dagger_i\rangle$. We will see in subsection 4.4 how to calculate
these densities in r-RPA. For the 1p-2p and 2p-2p GF we give the
particular combinations which are directly relevant for the calculation of the effective
potential~:
\begin{eqnarray}
\sum_{123}\,V_{1,-2,-3}\,G_{\phi_2 \phi_3\,,\,\phi^\dagger_1}(E) &=&
V\,\int {d{\vec P}\over (2\pi)^d}\,2\,\Sigma(E,{\vec P})\,G(E, {\vec P})
\nonumber\\
\sum_{1234}\,V_{1, 2,-3,-4}\,G_{\phi_3 \phi_4\,,\,\phi^\dagger_1 \phi^\dagger_2}(E)
 & = & V\,\int {d{\vec P}\over (2\pi)^d}\,b\,\bigg[2\, I(E,{\vec P})
 \,+\,b
{I^2(E,{\vec P})\over 1\,-\,{b\over 2}\,I(E,{\vec P})}\nonumber\\
& & +\,{b^2\,s^2}\,
\left({I(E,{\vec P})\over 1\,-\,{b\over 2}\,I(E,{\vec P})}\right)^2\,
G(E, {\vec P})\bigg]~.\label{GF22}
\end{eqnarray}
The reader may check that these results have a very clear diagrammatic interpretation
(see figure \ref{diag}).
From these expressions, one can get the expectation value of the three-body 
Hamiltonian by taking the appropriate $t'\to t$ limit of the GF. 
\begin{equation}
\langle H_3\rangle/ V= {1\over 6}\,\int {d{\vec P}\over (2\pi)^d}\,
\int\,{i \,dE \over (2\pi)}\,e^{i\,E\,\eta^+}\,2\,\Sigma(E,{\vec P})\,G(E, {\vec P})
\label{H3}~.
\end{equation}
The correlated part of the four-body Hamiltonian is obtained with  the same technique 
but the
uncorrelated GF has to be removed. As we will see below the expression of the
correlated energy will involve the following quantity~:
\begin{eqnarray}  
& &{1\over 24 V}\,\int\,{i \,dE\over (2\pi)}\,e^{i\,E\,\eta^+}\,\bigg[
\sum_{1234}\,V_{1, 2,-3,-4}\,\big(G_{\phi_3 \phi_4\,,\,\phi^\dagger_1
\phi^\dagger_2}(E)\,-\,
G^0_{\phi_3 \phi_4\,,\,\phi^\dagger_1\phi^\dagger_2}\bigg)\bigg]\nonumber\\ 
& &={1\over 24}
\int {d{\vec P}\over (2\pi)^d}\,
\int\,{i \,dE\over (2\pi)}\,e^{i\,E\,\eta^+}\,\bigg[
b^2 \,{I^2(E,{\vec P})\over 1\,-\,{b\over 2}\,I(E,{\vec P})}
\nonumber\\
& &\quad +\,{b^3\,s^2}\,
\left({I(E,{\vec P})\over 1\,-\,{b\over 2}\,I(E,{\vec P})}\right)^2\,
G(E, {\vec P})\bigg]\label{RESF}
\end{eqnarray}
where the first order term ({\it i.e.} the one loop term $I(E, {\vec P})$ of eq.
(\ref{GF22})) has been removed.

\subsection{The Effective Potential}

The starting Hamiltonian eqs. (\ref{HSTART}, \ref{H3H4}) 
contains a free Hamiltonian of bare particles with energy
${\cal O}_{\vec q}$ and an interacting Hamiltonian $H_3+H_4$. To simplify the writing 
we now
replace the momentum labels ${\vec q}$ by integers $i$ and omit the linear term 
in $\phi$ which does not directly play a role in the formal manipulations~:
\begin{equation}
H =  V\left({1\over 2} \mu^2_0\,s^2\,+\,{b\over 24}\,s^4\right)\,+\,
\sum_1\,{1\over 2}\left(\Pi_1\,\Pi^\dagger_1\,+\,{\cal O}^2_1\,
\phi_1\,\phi^\dagger_1\right)\,+\,H_3\,+\,H_4 .\label{HAM}
\end{equation}
In the RPA approach, the contribution of the interacting part 
of the Hamiltonian 
systematically  transforms,  in the expressions of the various double
commutators, the bare single particle energy 
into the generalized mean-field 
single particle energy (\ref{GGAP}) 
which is finite after mass renormalization in one spatial dimension~:
\begin{equation}
{\cal O}^2_1\,\to \,\varepsilon^2_1= {\cal O}^2_1\,+\,{b\over 2}
\,\langle \phi^2\rangle_R
\end{equation}
where $\langle\phi^2\rangle_R=(1/V)\sum \langle\phi_1\,\phi^\dagger_1\rangle$ 
is in principle the correlated  vacuum density (see subsection 4.4) except in standard RPA 
where it is taken on the Gaussian ground state.

\noindent
This suggests to rewrite the Hamiltonian as~:
\begin{equation}
H\,=\,H_0 \,+\,H_{int}
\end {equation}
with~: 
\begin{equation}
H_{int}=H_3\,+\,H_4\,-\,{b\over 4}\,\langle\phi^2\rangle_R\,\sum_1\,
:\phi_1\,\phi^\dagger_1:_\varepsilon\,-\,V\,{b\over 8}\,
\langle\phi^2\rangle_\varepsilon^2~.\label{HINT}
\end{equation}
$\langle\phi^2\rangle_\varepsilon$ is the expectation value taken on the ground state of the
mean-field quasi-particles with energies $\varepsilon_1$ and  $:\,...\,:_\varepsilon$ 
is the normal ordering with respect to this vacuum, namely~:
\begin{equation}
:\phi_1\,\phi^\dagger_1:_\varepsilon=\phi_1\,\phi^\dagger_1-
\langle\phi_1\,\phi^\dagger_1\rangle_\varepsilon~.
\end{equation}
We use again the notation $\langle\phi^2\rangle_R$ for the scalar density. 
In renormalized
RPA this scalar density is in principle the correlated one.
In standard RPA the quantity $\langle\phi^2\rangle_R$ appearing in $H_{int}$
is identified with the mean-field scalar density $\langle\phi^2\rangle_\varepsilon$ and
$\varepsilon$ refers to the Gaussian HFB mean-field. In that case, the
interacting Hamiltonian reduces to~:
\begin{equation}
(H_{int})_{standard\, RPA}\,=\,H_3\,+\,:H_4:_\varepsilon\, .\label{NORRPA}
\end{equation}
\noindent
The $H_0$ Hamiltonian is obtained as $H-H_{int}$~: 
\begin{eqnarray}
H_0 & =& V\left({1\over 2} \mu^2_0\,s^2\,+\,{b\over 24}\,s^4\right)\,+\,
\sum_1\,{1\over 2}\left(\Pi_1\,\Pi^\dagger_1\,+\,{\cal O}^2_1\,
\phi_1\,\phi^\dagger_1\right)\,+\nonumber\\
& & +\,{b\over 4}\,\langle\phi^2\rangle_R\,\sum_1\,
:\phi_1\,\phi^\dagger_1:_\varepsilon\,+\,V\,{b\over 8}\,
\langle\phi^2\rangle_\varepsilon^2~.
\end{eqnarray}
Again in the case of standard RPA the self-consistent scalar-density
$\langle\phi^2\rangle_R$
is replaced  in the above expression by the
Gaussian HFB scalar density $\langle\phi^2\rangle_\varepsilon$.
$H_0$ can be rewritten as~:
\begin{equation}
H_0=E_0\,+\,\sum_1\,{1\over 2}\left(:\Pi_1\,\Pi_1^\dagger:_\varepsilon\,+\,
\varepsilon^2_1\,:\phi_1\,\phi^\dagger_1:_\varepsilon\right)~.\label{H0}
\end{equation}
It has a form of a free Hamiltonian for quasi-particles with mass $m$ 
(see eq. (\ref{GGAP})) {\it i.e.} 
$m^2=\varepsilon_{\vec q}^2\,-\,{\vec q}^2$. In one spatial dimension, 
this mass is rendered finite by a simple mass renormalization~:
\begin{equation}
m^2=\mu^2\,+\,{b\over 2}\,s^2\,+\,{b\over 2}\,\left(\langle\phi^2\rangle_R
\,-\,\int_{-\Lambda}^{+\Lambda}\,{dq\over 2\pi}\,
{1\over 2\sqrt{q^2+\mu^2}}\right)~.
\end{equation}
$E_0$ is the generalized mean-field vacuum energy~:
\begin{eqnarray}
{E_0\over V} &=& {1\over 2} \mu^2_0\,s^2\,+\,{b\over 24}\,s^4\,+\,
{b\over 8}\,\langle\phi^2\rangle_\varepsilon^2\nonumber\\ 
& & +\,\sum_1\,{1\over 2}\left(\langle\Pi_1\,\Pi^\dagger_1\rangle_\varepsilon\,+\,
{\cal O}^2_1\,\langle\phi_1\,\phi^\dagger_1\rangle_\varepsilon\right)\nonumber\\
& = & {1\over 2} \mu^2_0\,s^2\,+\,{b\over 24}\,s^4\,+\,
{1\over 2 V}\,\sum_{\vec q}\,\left({\varepsilon_q\over 2}\,+\,{{\cal O}^2_q\over
2\varepsilon_q}\right)\, +{b\over 8}\,\langle\phi^2\rangle_\varepsilon\nonumber\\
&=& \mu^2\,\bigg[{1\over 2}\,s^2\,+\,p\,s^4\,+\,
{1\over 8\pi}\,\left({m^2\over \mu^2}\,-1\,-{m^2\over \mu^2}\,
\ln{m^2\over \mu^2}\right)\,-\,{3\,p\over 16\pi^2}\,
\left(\ln{m^2\over \mu^2}\right)^2\bigg]~.
\end{eqnarray}
We may notice that $E_0/V$ is formally equal to the Gaussian energy density 
(\ref{GAE1}, \ref{GAE2}). However the
single-particle energy $\varepsilon_{q}$ and the corresponding quasi-particle mass $m$
entering its expression now depends on the correlated scalar density $\langle\phi^2\rangle$,
in the self-consistent version.

\medskip\noindent
$\langle H_3\rangle$ and $\langle H_4\rangle$ can be calculated once the RPA 2p-1p and
2p-2p GF are known (see the end of the previous subsection). But to calculate  the
total energy we also need the expectation value on the RPA ground-state of the one-body
operators $\phi_{\vec q}\phi^\dagger_{\vec q}$ and 
$\Pi_{\vec q}\Pi^\dagger_{\vec q}$ which are not directly given by the RPA calculation.
This is the  well-known difficulty of RPA, even in its simplest standard form,
 which frequently appears in the context of nuclear physics. 
In other words the calculation of the kinetic energy in 
RPA needs further manipulations. One possible way to achieve this 
is to use the so-called charging formula \cite{FW} 
for the calculation of the correlation energy ({\it i.e.} the deviation from the
mean-field energy $E_0$) which has been historically introduced for the electron gas
problem. Here we will show how to adapt the charging formula beyond the standard RPA, namely
in the r-RPA case in the context of a quantum field theory.

\medskip\noindent
The idea is to introduce a Hamiltonian where the coupling constant is varying
between zero and its physical value. We thus define the auxiliary Hamiltonian~:
\begin{equation}
H'(\rho)\,=\,H_0\,+\,\rho\,H_{int}\, ,\qquad H'(\rho=1)=H~.
\end{equation}
The first thing to do is to solve the RPA problem for the $H'(\rho)$ Hamiltonian. 
For this purpose, one can notice that its explicit form is given by~:
\begin{eqnarray}  
H'(\rho)&=& V\left({1\over 2} \mu^2_0\,s^2\,+\,{b\over 24}\,s^4\right)\,+\,
\sum_1\,{1\over 2}\left(\Pi_1\,\Pi^\dagger_1\,+\,{\cal O}^2_1\,
\phi_1\,\phi^\dagger_1\right)\,+\,\rho(H_3\,+\,H_4)\nonumber\\
& & +\,(1\,-\,\rho)\,\left({b\over 4}\,\langle\phi^2\rangle_R\,\sum_1\,
:\phi_1\,\phi^\dagger_1:_\varepsilon\,+\,V\,{b\over 8}\,
\langle\phi^2\rangle_\varepsilon^2\right)~.
\end{eqnarray}
Up to constant terms, the Hamiltonian $H'(\rho)$ can be rewritten as~:
\begin{eqnarray}  
H'(\rho)&=& V\left({1\over 2} \mu^2_0\,s^2\,+\,{b\over 24}\,s^4\right)\,+\,
\sum_1\,\bigg[{1\over 2}\,\Pi_1\,\Pi_1^\dagger\nonumber\\ 
& &+{1\over 2}\,
\left({\cal O}^2_1\,+\,{b\over 2}\,(1\,-\,\rho)\,\langle\phi^2\rangle_R\right)\,
\phi_1\,\phi^\dagger_1\bigg]\,+\,\rho\,(H_3\,+\,H_4)
\end{eqnarray}
For what concerns the solution of the $H'(\rho)$ RPA problem (in practice for the
calculation of the commutators and 
double commutators entering the RPA equations) one has to make the following modifications 
with respect to the $H$ problem~:
\begin{eqnarray}
H\,&\to & \,H'(\rho)\nonumber\\
{\cal O}^2_1 &\to &\,{\cal O}^2_{1\rho}=  {\cal O}^2_1\,+\,{b\over 2}\,(1\,-\,\rho)\,
\langle\phi^2\rangle_R \nonumber\\
H_3+H_4\,&\to &\,\rho\,(H_3\,+\,H_4) 
\end{eqnarray}
The single-particle energy occurring in the self-consistent RPA GF will 
be thus modified according to~:
\begin{eqnarray}
\varepsilon^2_1\,\to \,\varepsilon^2_{1\rho} &=&
{\cal O}^2_{1\rho}\,+\,{b\over 2}\,\rho\,\langle\phi^2\rangle_{R\rho} \nonumber\\
&=&{\cal O}^2_1\,+\,{b\over 2}\,\left( (1-\rho)\langle\phi^2\rangle_R\,+\,
\rho\,\langle\phi^2\rangle_{R\rho}\right)\nonumber\\
&=& \varepsilon^2_1\,+\,{b\over 2}\,\rho\,\left(\langle\phi^2\rangle_{R\rho}\,-\,
\langle\phi^2\rangle_R\right)\label{MOD}
\end{eqnarray}
where $\langle\phi^2\rangle_\rho$ is the scalar density in the correlated RPA ground state of 
$H'(\rho)$.
Again the notation $\langle\phi^2\rangle_{R\rho}$ is employed~: in r-RPA, it coincides with
the scalar density calculated on the self-consistent ground-state of the $H'(\rho)$
Hamiltonian, while in the standard RPA, it coincides with the Gaussian density, i.e.
calculated on the ground state of $H_0$.  
The solution of the $H'(\rho)$ r-RPA problem is obtained formally from the solution of
the $H$ r-RPA problem (eq.  \ref{RESI}-\ref{RESF}) by simply replacing $\varepsilon_1$ by 
$\varepsilon_{1\rho}$, the coupling constant $b$ by $\rho\,b$ and ${\cal
N}_1=\langle\phi_1\phi_1^\dagger\rangle_R$ by  
${\cal N}_{1\rho}=\langle \phi_1\phi_1^\dagger\rangle_{R\rho}$ calculated
self-consistently with the $H'(\rho)$ Hamiltonian. 

\smallskip\noindent
In the standard RPA all the expectation values of $\phi^2$ 
are taken on the Gaussian ground state. In
this case the energies $\varepsilon_\rho$ remain identical to the Gaussian single-particle
energies $\varepsilon$~:

$\hbox {standard RPA~:}\quad \langle\phi^2\rangle_{R\rho}=\langle\phi^2\rangle_R=
\langle\phi^2\rangle_\varepsilon\, ,
\qquad \varepsilon_{1\rho}=\varepsilon_1~.$

\medskip\noindent 
Once the r-RPA problem is solved one can calculate the RPA ground state energy 
relative to the starting Hamiltonian.  Since both $H_0$ and $H_{int}$ are independent 
of $\rho$ and since $H'(\rho=1)$ coincides with the original $H$ one can apply the
charging formula. The RPA ground state energy can be obtained as~:
\begin{equation}
E_{RPA}=E_0\,+\,\int_0^1\, {d\rho\over\rho}\, \langle \rho\,H_{int}\rangle_\rho
\label{ETOT}\end{equation}
where $E_0$ is the already calculated generalized mean-field energy. Using Wick theorem with respect 
to the vacuum of quasi-particles with energies $\varepsilon_\rho$, the correlated part can
be rewritten as~:
\begin{eqnarray}
\langle \rho\,H_{int}\rangle_\rho &=& 
\langle \rho \,H_3\rangle_\rho\,+\, \langle \rho\, :H_4:_{\varepsilon_\rho}\rangle_\rho
-\,V\,{\rho\,b\over 8}\,\left(\langle \phi^2\rangle_{\varepsilon_\rho}\,-\,
\langle \phi^2\rangle_\varepsilon\right)^2\nonumber\\
& &-\,V\,{\rho\,b\over 4}\,\left(\langle \phi^2\rangle_R\,-\,
\langle \phi^2\rangle_{\varepsilon_\rho}\right)
\left({1\over V}\sum_1\,\langle \phi_1\phi_1^\dagger\rangle_\rho\,-\,
\langle \phi^2\rangle_\varepsilon\right)~.\label{HINT2}
\end{eqnarray}

\smallskip\noindent
In this formula $\langle\phi^2\rangle_R$ is as before 
the self-consistent scalar density of the
original $H$. $\langle\phi^2\rangle_{\varepsilon}$ is the scalar density on the
generalized mean field vacuum (vacuum of quasi-particles with energy
$\varepsilon_{\vec q}$) in the $H$ problem and 
$\langle\phi^2\rangle_{\varepsilon_\rho}$ corresponds to the equivalent quantity for the
$H'(\rho)$ Hamiltonian. The remaining expectation values
noted $\langle\,..\,\rangle_\rho$  have to be taken on the r-RPA
ground-state of $H'(\rho)$. The calculation of these latter expectation values are
made using eq.({\ref{RESI}-\ref{RESF}) where all the quantities are now relative to the
$H'(\rho)$ problem as explained before. 

\noindent
In the particular case of the standard RPA the extra term in the expression of the correlation
energy disappears since, following eq.(\ref{NORRPA}), one has~:  
 \begin{eqnarray}
\langle \rho\,H_{int}\rangle_\rho &=& 
\langle \rho \,H_3\rangle_\rho\,+\, \langle \rho\, :H_4:_{\varepsilon}\rangle_\rho ~.
 \end{eqnarray} 
As we will see explicitly in the next subsection the expectation value of the normal ordered
Hamiltonian will involve an integration over the calculated Green's functions. In the case 
of the standard RPA the $\rho$ integration can be done analytically. This is not the case 
in the r-RPA since these Green's functions will involve the $\varepsilon_\rho$'s and  the
self-consistent densities ${\cal N}_\rho$ which depend explicitly on $\rho$.

\subsection{Results in standard RPA in 1+1 dimension}

\noindent
{\it Single particle mode}

\smallskip\noindent
The RPA single particle mode $\omega_P$ 
 with momentum ${\vec P}$ is obtained as the solution of
the equation see {\it e.g.} eq.(\ref{RESI})~:
\begin{equation}
\omega^2_P=\varepsilon^2_P\,+\,\Sigma(E=\omega_P,\, {\vec P})~.
\end{equation}
In the standard RPA, the densities are simply taken as ${\cal N}_1=1/2\varepsilon_1$ where
the $\varepsilon_1$'s are the Gaussian single-particle energies.  In that case 
$I(E, {\vec P})$ (eq. \ref{IEP}) and consequently $\Sigma(E, {\vec P})$ (eq. \ref{SEP})
are explicitly covariant in the
sense that they depend only on $E^2\,-\,{\vec P}^2$ and not 
on $E$ and ${\vec P}$ separately.
After a simple boost-like change of variables one can show that~:
\begin{eqnarray}  
I(E, {\vec P}) &\equiv & I(E^2\,-\,{\vec P}^2)\nonumber\\
&=& \int \,{d{\vec t}\over (2\pi)^d}\,{1\over \varepsilon_t}\,
{1\over E^2\,-\,{\vec P}^2\,-\,4\,\varepsilon_t^2\,+\, i\eta}~.
\end{eqnarray}
Consequently the RPA mode has a dispersion relation which is $\omega^2_P= M^2\,+\,
{\vec P}^2$. The mass $M$ of the single-particle RPA mode is thus the solution of the
equation~:
\begin{equation}
M^2\,=\,m^2\,+\,\Sigma(E^2\,-\,{\vec P}^2=M^2)~.
\end{equation}
In one spatial dimension the RPA mass operator is finite and there is no need of
further coupling constant renormalization. In figure \ref{normalMass}
the result of the calculation for $M$ in one spatial dimension is shown for various values 
of the dimensionless
coupling constant $p$ as a function of $s$. It is apparent that for $p$ larger than a
certain value the RPA equation may have an imaginary solution. 
Such a feature, which can appear
 in RPA, simply means that the HFB ground state is unstable. Hence for that
particular theory one has to go to a superior version of the HFB-RPA approach. 

\begin{figure}
\begin{center}
\epsfysize=8 true cm \epsfxsize=10true cm
\epsffile{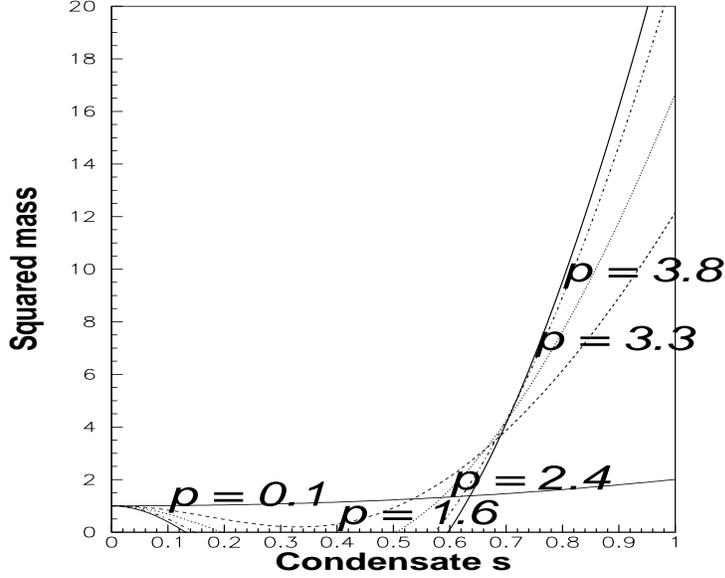} 
\caption{Squared standard RPA mass (in unit of $\mu^2$) as a function of $s$.} 
\label{normalMass} 
\end{center}
\end{figure}

\smallskip\noindent
{\it Correlation energy}

\smallskip\noindent
Although the standard RPA leads, in this particular theory, to an instability 
in a certain range of coupling constants, 
it is however
interesting to look at the expression of the correlation energy at least at a formal
level. We divide the correlation energy density  in three pieces
\begin{equation}
{E_{corr}\over V}= {E^{(3)}_{corr}\over V}\,+\,{E^{(4c)}_{corr}\over V}\,+\,
 {E^{(4nc)}_{corr}\over V}~.
 \end{equation}
According to eqs.(\ref{H3}, \ref{ETOT}), the piece corresponding to the three-body 
Hamiltonian writes~: 
 \begin{eqnarray}
{E^{(3)}_{corr}\over V}&=& \int_0^1{d\rho\over\rho}\,\int {d{\vec P}\over (2\pi)^d }\,
\int\,{i \,dE\over (2\pi)}\,e^{i\,E\,\eta^+}\,{\rho^2\,b^2\,s^2\over 6} \,
{I(E,{\vec P})\over 1\,-\,{\rho\,b\over 2}\,I(E,{\vec P})}\,\times\nonumber\\ 
& &
\left(E^2\,-{\vec P}^2\,-\,m^2\,-\,
{\rho^2\,b^2\,s^2\over 2} \,{I(E,{\vec P})\over 1\,-
\,{\rho\,b\over 2}\,I(E,{\vec P})}\right)^{-1}~.\label{E3C}
\end{eqnarray} 
Notice that, as mentioned before, the $\rho$ integration can be performed analytically.  
It is convenient to transform the energy integration into an integration on the 
imaginary axis (Wick rotation) by making the change of variable $E=i\,z$. In the usual 
RPA, the loop $I(E, {\vec P})$ and thus the whole integrand $h_3$ actually depends only on
$E^2-{\vec P}^2$. After the Wick rotation it depends only on ${\cal S}=z^2+{\vec P}^2$ 
and the momentum integration can be done analytically in 1+1 dimension~:
\begin{equation}
\int_{-\infty}^{+\infty}\, {dP\over 2\pi}\,
\int\,{i \,dE\over (2\pi)}\,e^{i\,E\,\eta^+}\, h_3(E^2\,-\, {\vec P}^2)
=-\int_0^\infty\,{d{\cal S}\over 4\pi}\,h_3(-{\cal S})~.\label{TRICK}
\end{equation}

\smallskip\noindent
For what concerns the four-body interacting piece we start from the explicit form of its
normal ordering with respect to the $\varepsilon$ basis~:
\begin{eqnarray}
\langle:H_4:_\varepsilon\rangle &= &{b\over 24\, V\,\sqrt{\prod_i 2\varepsilon_i}}
\,\delta_{1+2+3+4}\bigg(
\langle b^\dagger_1\, b^\dagger_2\,b^\dagger_3\,b^\dagger_4\,
+\, b_{-1}\, b_{-2}\,b_{-3}\,b_{-4}\,\nonumber\\
& & +\,4\,b^\dagger_1\, b^\dagger_2\,b^\dagger_3\,b_{-4}
\,+\,4\,b^\dagger_1\,b_{-2}\, b_{-3}\,b_{-4}\,
\,+\,6\,b^\dagger_1\, b^\dagger_2\,b_{-3}\,b_{-4}\rangle\bigg)
\end{eqnarray}
where summation over repeated indices is now omitted. 
As explained in subsection 4.1, these matrix elements can be evaluated as an energy
integral of two-particle Green's functions whose explicit expressions are given 
in the appendix. Noticing  that all the matrix elements 
$\langle b^\dagger b^\dagger  b^\dagger b\rangle$ are identically zero,
it is convenient, after standard manipulations, 
to split $ \langle:H_4:_\varepsilon\rangle$ into two pieces~: 
\begin{eqnarray}
\langle:H_4:_\varepsilon\rangle &=&\langle :H_4:_\varepsilon\rangle^{(c)}\,+\,
\langle :H_4:_\varepsilon\rangle^{(nc)}\nonumber\\
\langle :H_4:_\varepsilon\rangle^{(c)} & =& {b\over 24\, V\,
\sqrt{\prod_i 2\varepsilon_i}}\,
\delta_{1+2+3+4}\bigg(
\langle ( b^\dagger_1\, b^\dagger_2\,+\,b_{-1}\, b_{-2})\,
(b_{-3}\,b_{-4}\,+\,b^\dagger_3\,b^\dagger_4)\rangle\,\nonumber\\
& & -\,
2\,\delta_{1+3}\,\delta_{2+4}\,(1\,+\,2\,\langle
b^\dagger_1\,b_1\rangle)\bigg)\nonumber\\
&=& {b\over 24\, V}\,\int\,{i \,dE\over (2\pi)}\,e^{i\,E\,\eta^+}\,\bigg[
\,\delta_{1+ 2-3-4}\,\big(G_{\phi_3 \phi_4\,,\,\phi^\dagger_1
\phi^\dagger_2}(E)\,-\,
G^0_{\phi_3 \phi_4\,,\,\phi^\dagger_1\phi^\dagger_2}\bigg)\bigg]\nonumber\\
\langle :H_4:_\varepsilon\rangle^{(nc)} & =& {b\over 24\, V\,
\sqrt{\prod_i 2\varepsilon_i}}\,4\,\langle b^\dagger_1\, b^\dagger_2\,b_{-3}
\,b_{-4}\rangle\nonumber\\
&=&{b\over 24\, V}\,\int\,{i \,dE\over (2\pi)}\,e^{i\,E\,\eta^+}\,\bigg[
\,\delta_{1+ 2-3-4}\,G_{b_3 b_4\,,\,b^\dagger_1
b^\dagger_2}(E)\bigg],
\end{eqnarray}
where the suffices $(c)$ and $(nc)$ stand for covariant and non covariant in a 
sense to be discussed just below. 
It is important to notice that the above result remains valid even in the case of the r-RPA 
where the occupation numbers $\langle b^\dagger b\rangle$ do not vanish. 
Using the results of the appendix and the charging formula, the corresponding
contributions to the correlation energy can now be obtained~:
\begin{equation}
{E^{(4c)}_{corr}\over V}= \int_0^1{d\rho\over\rho}\,\int {d{\vec P}\over (2\pi)^d }\,
\int\,{i \,dE\over (2\pi)}\,e^{i\,E\,\eta^+} \, 
I^2(E\, ,\,\vec P)\,F(E\, ,\,\vec P,\, \rho ) \label{E4C}
\end{equation}
\begin{equation}
{E^{(4nc)}_{corr}\over V}= \int_0^1{d\rho\over\rho}\,\int {d{\vec P}\over (2\pi)^d }\,
\int\,{i \,dE\over (2\pi)}\,e^{i\,E\,\eta^+} \, 
I^{(1)\,2}(E\, ,\,\vec P )\,F(E\, ,\,\vec P \, ,\,\rho)\label{E4NC}
 \end{equation}
where the $\rho$ integration can be again performed analytically, in standard RPA.
$F(E\, ,\,\vec P\, , \,\rho)$ has the explicit expression~:
\begin{eqnarray}
F(E\, ,\,\vec P\, , \,\rho) &\equiv & F(E^2\,-\,\vec P^2, \, \rho) =
{1\over 24}\bigg(
 \,{\rho^2\,b^2\over 1\,-\,{\rho\,b\over 2}\,I(E,{\vec P})}
\,+\,\,{\rho^3\,b^3\,s^2\over\big( 1\,-\,{\rho\,b\over 2}\,I(E,{\vec
P})\big)^2}\times\nonumber\\
& &\left(E^2\,-{\vec P}^2\,-\,m^2\,-
{\rho^2\,b^2\,s^2\over 2} \,{I(E,{\vec P})\over 1\,-
\,{\rho\,b\over 2}\,I(E,{\vec P})}\right)^{-1}\bigg)~.
\end{eqnarray}
The contribution $E_{corr}^{(4c)}$ is explicitly covariant in the sense that the integrand
depends only on $E^2-\vec P^2$ and the trick of eq.(\ref{TRICK}) can be applied again. 
For what
concerns $E_{corr}^{(4nc)}$, the explicit calculation is more delicate since 
\begin{equation}
I^{(1)}(E,{\vec P}) =
\int \,{d{\vec k}_1\,d{\vec k}_2\over (2\pi)^d}\,{\delta^{(d)}\left({\vec P}-
{\vec k}_1 - {\vec k}_2\right)\over
 2\,\varepsilon_1\,2\,\varepsilon_2}\,
{1\over 
E\,-\,\varepsilon_1\,+\,\varepsilon_2+i \eta},
\end{equation}
appearing in eq.({\ref{E4NC}),
depends separately on $E^2$ and $P^2$. However, as it is familiar in nuclear physics,
$E_{corr}^{(4nc)}$ which involves expectation values of the  $b^\dagger b^\dagger b b$'s 
is of higher order in the $Y$ amplitudes than $E_{corr}^{(4c)}$ involving $b b b b$ 
ground state matrix elements. Indeed, it can be checked analytically that to leading order
in the interaction ({\it i.e.} replacing $F(E, \vec P, \rho)$ by a constant value 
$\rho^2 b^2/24$), $E_{corr}^{(4nc)}$ identically vanishes.

\smallskip\noindent
Hence, we find that the correlation energy contains a piece, $E_{corr}^{(4nc)}$, 
which is manifestly non covariant even in the standard RPA . 
This problem has not been pointed out before, since, to our
knowledge, the RPA correlation energy has never been calculated in a case of a
relativistic theory for bosons.  We have neglected this contribution
in our preliminary numerical estimate for the reasons given just above. Nevertheless we give the explicit result, 
involving a four-dimensional integration, of this non covariant contribution. After some
manipulations and change of integration variables, one obtains~: 
\begin{equation}
{E^{(4nc)}_{corr}\over V}=\int_0^1{d\rho\over\rho}\,
\int_0^\infty\,{d{\cal S}\over 2\pi^2}\,\int_0^{\pi/2}\,d\theta
\bigg({\cal S}\, \cos^2\theta\,J^2({\cal S}\,,\,\theta)\,-\,I^2(-{\cal S})
\bigg)\,F(-{\cal S})\label{TRICKNC}
\end{equation}
with ~:
\begin{equation}
J({\cal S}\,,\,\theta)=-\int \,{d t\over 2\pi}\,
{1\over \varepsilon_t\,\sqrt{4 \varepsilon_t^2\,+{\cal S} \sin^2\theta}}\,
{1\over {\cal S}\,+\,4\,\varepsilon_t^2}\label{JINT}
\end{equation}
to be compared with~:
\begin{equation}
I(-{\cal S})=-\int \,{d t\over 2\pi}\,
{1\over \varepsilon_t}\,
{1\over {\cal S}\,+\,4\,\varepsilon_t^2}~.\label{IINT}
\end{equation}

\begin{figure}
\begin{center}
\epsfysize=4 true cm \epsfxsize=13 true cm
\epsffile{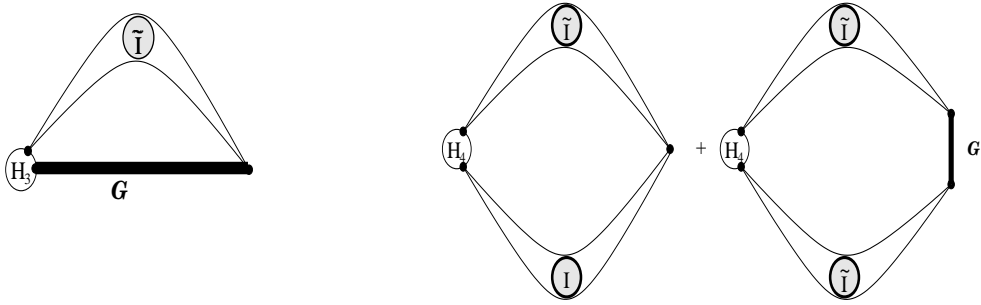}  
\caption{Diagrammatic view of the correlation energy. In the actual calculation, 
the RPA one-particle propagator (thick line) is replaced by the mean-field one.}
\label{diag}
\end{center}
\end{figure}

\smallskip\noindent
As mentioned above there is an instability of the HFB ground state against RPA
fluctuations, which makes the 
correlation energy divergent. However, to have a first idea of the influence of the RPA 
fluctuations  we replace in the above expressions for the correlation energy 
 the RPA one-particle propagator 
by the mean field one. This is illustrated in figure \ref{diag} where a diagrammatic 
interpretation of the RPA correlation energy is shown. Adding this correlation energy to the
mean-field energy $E_0$ one obtains the RPA effective potential {\it i.e.} 
the RPA energy versus the
condensate $s$ for various values of the dimensionless coupling constant $p$.
One gets a second-order phase transition with a
critical coupling $p_c=1.8$ (see figure \ref{nrpaen}). 
This has to be compared with the lattice result \cite{LW97} and cluster expansion technique 
\cite{HCPT95} showing a second order transition respectively at 
$p_c=2.55$ and $p_c=2.45$. It is fair to mention that the neglected non covariant contribution 
$E^{4c}_{corr}$ is repulsive and will likely push the critical coupling constant
to a higher value closer to the lattice result. 
Although this result is encouraging, it is obviously
needed to go beyond the standard RPA to eliminate the unphysical instability mentioned 
above and seen in fig. 3.

\begin{figure}
  \begin{minipage}[b]{0.48\linewidth}
    \centering\epsfig{figure=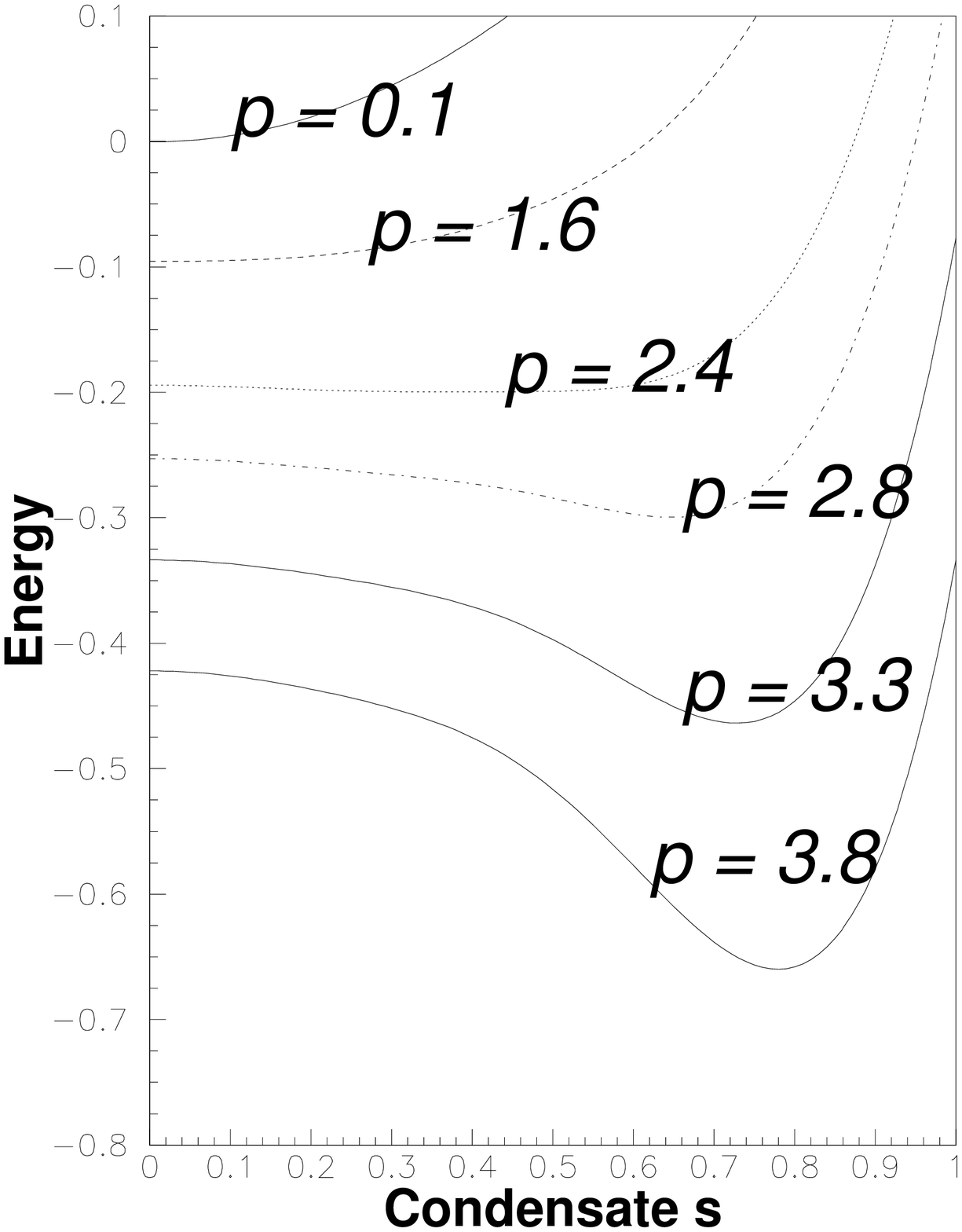,width=\linewidth}
  \end{minipage}\hfill
   \begin{minipage}[b]{0.48\linewidth}
    \centering\epsfig{figure=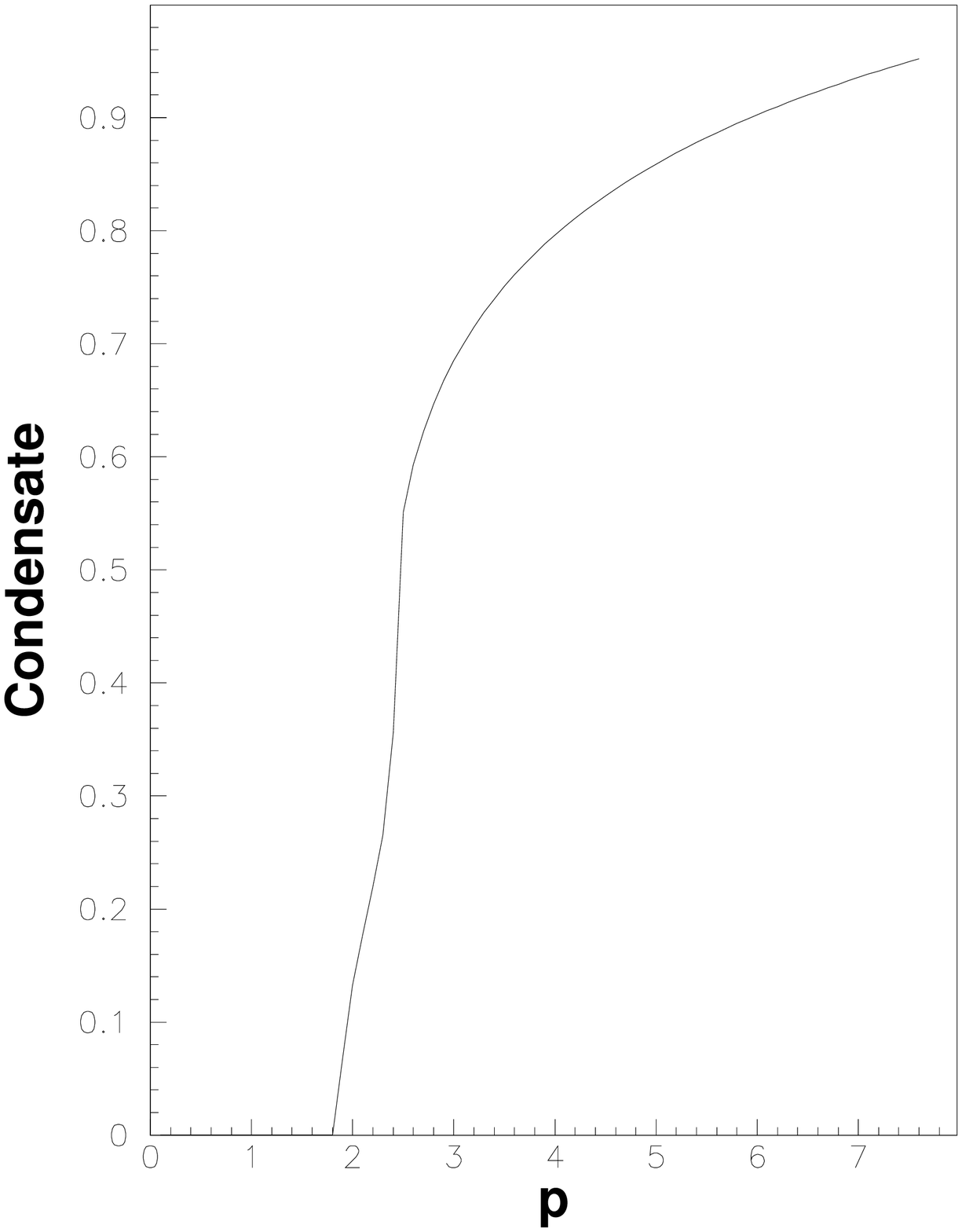,width=\linewidth}
  \end{minipage}
%\par\vspace*{-30mm}
\caption{
 Left panel~: Effective potential for various values of the
dimensionless coupling constant $p$ in standard RPA. Right panel~; 
value of the condensate minimizing
the standard RPA effective potential as a function of $p$.}
\label{nrpaen}
\end{figure}

\subsection{Results in renormalized  RPA in 1+1 dimension}

\noindent
{\it Single particle mode}

\smallskip\noindent
In view of the calculation of the correlation energy we have now to solve the RPA problem 
for any value of $\rho$. The main problem is thus to determine the self-consistent 
${\cal N}_{{\vec q},\rho}\equiv\langle \phi_{\vec q}\phi^\dagger_{\vec q}\rangle_\rho$.
One very usual possibility is to calculate it self-consistently according to~:
\begin{equation}
{\cal N}_{{\vec P} \rho}=\int {i dE\over 2\pi}\,e^{i E \eta^+}\, G_\rho(E, {\vec P})
\label{SPEC}\end{equation}
where $G_\rho(E, {\vec P})$ is the one-particle propagator for the $H'(\rho)$ problem. 
One serious difficulty is that covariance is now lost in the sense that the loop integral 
$I_\rho(E, {\vec P})$ and consequently the mass operator $\Sigma_\rho(E, {\vec P})$
 depends separately on $E$ and ${\vec P}$ due to the presence of the density ${\cal N}$ 
in its
expression. This is certainly a weakness of the present approach. 
However, as discussed above, even standard RPA seems to have problems with covariance 
so we think that this additional difficulty simply reflects the fact that there is a
general problem of RPA with respect to covariance. Further work is needed to clarify
this point. On the other hand, one natural possibility  to recover covariance consists in 
imposing that the correct $I_\rho(E, {\vec P})$
is obtained through its CM expression according to~: 
\begin{eqnarray}  
I_\rho(E, {\vec P}) &\equiv & I_\rho(E^2\,-\,{\vec P}^2)\nonumber\\
&=& \int \,{d t\over 2\pi}\,
{2{\cal N}_{t \rho }\over E^2\,-\,{\vec P}^2\,-\,4\,\varepsilon_{t \rho }^2\,+\, i\eta}~.
\end{eqnarray} 
Lets us call $\Omega_{P \rho }=\sqrt{M^2_\rho \,+\,P^2}$ the RPA single particle mode which
is solution of the equation~:
\begin{equation}
\Omega^2_{ P\rho}=\varepsilon^2_{P\rho }\,+\,\Sigma_\rho(E^2\,-\,{\vec P}^2=M^2_\rho)~.
\label{MODE}
\end{equation}
In the quasi-particle approximation, the solution for ${\cal N}_{ P \rho }$ is~:
\begin{equation}
{\cal N}_{P \rho }\equiv\langle \phi_{\vec P}\phi^\dagger_{\vec P}\rangle_\rho=
{1\over 2\Omega_{P \rho}}~.
\end{equation}
The self-consistent equation  for the density thus becomes an equation for the mass of the
RPA single-particle mode which explicitly writes~:
\begin{equation}
M^2_\rho= m^2_\rho\,+\,{\rho^2\,b^2\,s^2\over 2}\,\left({I_\rho\over 1\,-\,{\rho\,b\over
2}I_\rho}\right)(E^2-P^2=M^2_\rho)
\end{equation}
with the generalized mean-field single-particle mass given by~: 
\begin{equation}
m^2_\rho=\mu^2\,+\,{b\over 2}\,s^2\,+\,{b\over 2}\,\left(\langle\phi^2\rangle
\,-\,\int_{-\Lambda}^{+\Lambda}\,{dq\over 2\pi}\,
{1\over 2\sqrt{q^2+\mu^2}}\right)\,+\,{\rho\,b\over
2}\,\left(\langle\phi^2\rangle_\rho\,-\,\langle\phi^2\rangle\right)
\end{equation}
which follows directly from (\ref{MOD}).  Hence we see that the equations  for the 
generalized mean-field energy and for the single particle RPA mode are now 
coupled, due to the presence of the self-consistent scalar density 
$\langle \phi^2\rangle$.  They finally reduce to determine $M_\rho$.
The procedure to solve the resulting equation at a given 
value of $b$ and $s$ is the following. 
We first solve the equation for $\rho=1$, which gives the RPA mode mass $M$, 
the densities ${\cal N}_P$,  $\langle\phi^2\rangle=\sum_P {\cal N}_P/V$
and  the mean-field mass $m$.  Once this is done we solve for $M_\rho$ which allows to 
obtain $m_\rho$ and $\varepsilon_{\rho P}=\sqrt{P^2+m^2_\rho}$.

\medskip\noindent
We show on figure \ref{SquaredMass} the results of the calculation for the RPA mass $M$. 
We see that the
instability problem has now disappeared. This is a first important success of 
the renormalized RPA.

\smallskip\noindent
{\it Correlation energy}

\smallskip\noindent
The calculation of the correlation energy can now be
done by assuming again covariance in the sense explained just above. 
The results of eqs.(\ref{E3C}, \ref{TRICK}, \ref{E4C}, \ref{E4NC}, \ref{TRICKNC}) 
can be applied by
just making in the final expressions, the replacements~:
\begin{equation}
I(-{\cal S})\,\to\,I_\rho(-{\cal S}) =-\int \,{d t\over 2\pi}\,
{2{\cal N}_{ t\rho}\over {\cal S}\,+\,4\,\varepsilon_{t\rho }^2}
\end{equation}
\begin{equation}
J({\cal S}\,,\,\theta)\,\to\,J_\rho({\cal S}\,,\,\theta) =-\int \,{d t\over 2\pi}\,
{1\over \sqrt{4 \varepsilon_{ t\rho}^2\,+{\cal S} \sin^2\theta}}\,
{2{\cal N}_{ t\rho}\over {\cal S}\,+\,4\,\varepsilon_{ t\rho}^2}~.
\end{equation}
In figure \ref{rrpaen}, we show the effective potential for various values of the
dimensionless coupling constant $p$, again neglecting the non covariant piece
(\ref{TRICKNC}) for reasons explained above. For $p$ below  $\simeq 2$ there is 
only one minimum 
at $s=0$ {\it i.e.} corresponding to a  symmetry unbroken phase. Beyond this value, 
a weakly pronounced minimum 
starts to develop at finite $s$. The symmetry broken phase becomes stable at $p_c\simeq 2.3$
indicating a  very weak first order 
transition. It is satisfying to see that the value of $p_c$ has moved in the right direction
towards the value given by cluster expansion \cite{HCPT95} 
and lattice calculation \cite{LW97}. 
It remains to calculate the non covariant contribution to see
if it is able to transform this weak first order transition into a genuine second order one.
Work in this direction is now in progress.

\subsection{Towards full renormalized RPA}

In the previous section we have presented a version of r-RPA which incorporates RPA
correlations in the scalar density which are induced  by the presence of a non-vanishing
condensate. However in the usual non symmetry broken  case this version is still equivalent 
to the  standard RPA. To go beyond standard RPA is, as already stated,  a difficult
problem. One possibility is to introduce the dynamical mass operator 
modifying the single particle propagator. This mass operator is of the form~:
\begin{eqnarray}
& &\Sigma^{(d)}_\alpha(t, t') = 2\,\varepsilon_\alpha\,\langle -i\,T\big( [H, b_\alpha](t) , 
[H, b^\dagger_\alpha](t')\big)\rangle_{irr}\nonumber\\
&=& {1\over 36}\,\sum_{1 23}\sum_{1'2'3'}V_{\alpha,-1,-2,-3}
\langle -i\,T\big( \phi_1 \phi_2 \phi_3(t)\, ,\, 
\phi_1'^\dagger \phi_2'^\dagger \phi_3'^\dagger(t')\big)\rangle
V_{\alpha,-1',-2',-3'}~.
\end{eqnarray}
Applying perturbation theory on top of the already calculated one particle propagator, one obtains for
the full propagator~: 
\begin{equation}
G_\alpha^{RPA}(t,t')=G_\alpha (t,t')\,+\int dt_1\,dt_2\,G_\alpha (t,t_1)
\Sigma^{(d)}_\alpha(t_1, t_2)\,G_\alpha (t_2,t')~.
\end{equation}
Using a factorization approximation, one obtains~: 
\begin{eqnarray}
\Sigma^{(d)}_\alpha(t, t') &=& {1\over 6}\,\sum_{1 23}\sum_{1'2'3'}
V_{\alpha,-1,-2,-3}\times\nonumber\\
& & \delta_{1,1'}\, G_1(t, t')\,G_{\phi_1\phi_2,\phi^\dagger_{2'}\phi^\dagger_{3'}}(t, t')\,
V_{\alpha,-1',-2',-3'}~.
\end{eqnarray}
The density ${\cal N}_\alpha$ can be in principle calculated as~:
\begin{equation}
{\cal N}_\alpha=\langle \phi^\dagger_\alpha\phi_\alpha\rangle 
\equiv i\, \lim_{t' \to t^+} G_\alpha^{RPA}(t,t')\equiv
\int {i\, dE\over 2\pi}\, e^{i E \eta^+}\, G_\alpha^{RPA}(E)~.
\end{equation}
The solution of this problem, {\it i.e.} to find self-consistently the scalar densities, 
is both formally and 
numerically very involved. However, as shown in a separate publication \cite{CDOS02} 
at least in the symmetry involved
region of the anharmonic oscillator, this procedure reproduces to leading order 
in $Y^2$ the correct 
occupation number from the exact SCRPA ground state wave function.  

\begin{figure}[p]
\begin{center}
\epsfysize=8 true cm \epsfxsize=10true cm
\epsffile{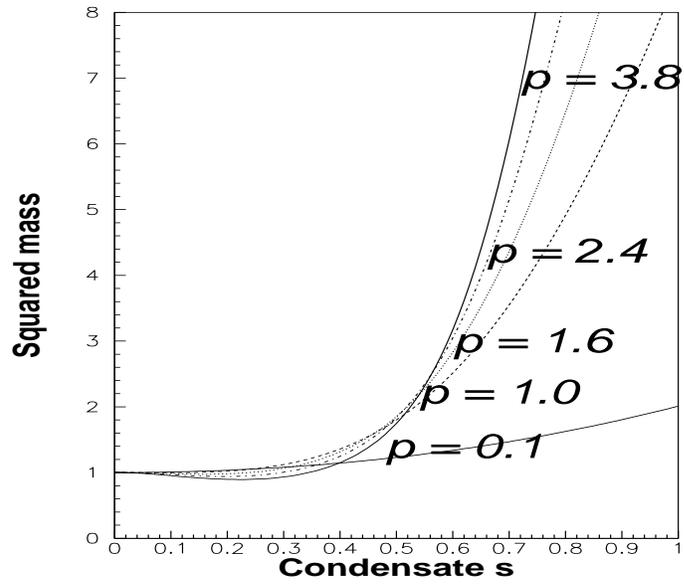}  
\caption{Squared r-RPA mass as a function of $s$.} 
\label{SquaredMass} 
\end{center}
\end{figure}

 \begin{figure}[p]
\begin{center}
\epsfysize=8true cm \epsfxsize=10true cm
\epsffile{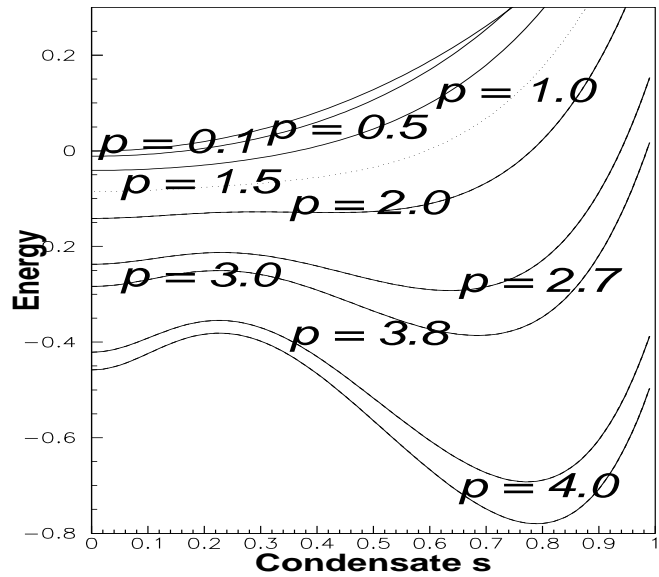}  
\caption{Effective potential for various values of $p$ in r-RPA.} 
\label{rrpaen}
\end{center}
\end{figure}

%%%%%%%%%%%%%%%%%%%%%%%%%%%%%%%%%%%%%%%%%%%%%%%%%%%%%%%%%%%%%%%%%%%%%%%%%%%%%%%
\section{Further remarks and discussions on the formalism.}
%%%%%%%%%%%%%%%%%%%%%%%%%%%%%%%%%%%%%%%%%%%%%%%%%%%%%%%%%%%%%%%%%%%%%%%%%%%%%%%

It is well known in non-relativistic many body theory that standard RPA
corresponds to a bosonization either of pairs of fermion operators or pairs
of boson operators. As a matter of fact, the RPA has first been invented
for Fermi systems and the bosonization of boson pair operators has
appeared much later in the literature \cite{curu,bijk}. However, in any case the
bosonization of boson pair operators (which are themselves NOT ideal
bosons) goes along the same lines as in the fermion case \cite{blai} and
also the generalization to relativistic field theory \cite{ASW97} presents no
particular obstacle (as a specific example, the bosonization technics has
extensively been studied in the case of the NJL model, see e.g. \cite{ripk}).
In doing so, one would naturally replace the boson
pair operators present in the definition of the Green's function (29) by
ideal boson operators, that is for example $ {b_q}^+{b_q'}^+  \rightarrow
{B_{qq'}}^+$, with $ {B_{qq'}}^+$ being an ideal boson operator. With such
a bosonization scheme the equation of motion for the Green's function
corresponding to \ref{DEA29} would lead to an inhomogeneity ${\cal N}_{ab}$ in
the Dyson equation of \ref{RPAEQC} which contains only unit operators on the
diagonal. Inspection of \ref{RPAEQC}, however, reveals that ${\cal N}_{ab}$
contains in addition one body expectation values. This fact stems from
the particular feature that in our approach we do not bosonise but stay
with the original boson pair operators. Contrary to the usual two body
Green's function technique, the boson pair operators are, however,
taken to be at equal times and not at two different  times. This means
that we single out  the S-channel. It is not widely known that an exact
integral equation for such two body-two time Green's functions can be
established analogous to the Dyson equation in the one-body case. The
mass operator of  this 'Two Body Dyson Equation' presents, as in the one
body case, an instantaneous mean field type part and a truly dynamical
part depending on the energy. Neglecting the latter leads precisely to
the selfconsistent equation \ref{RPAEQC}. It has turned out that in this way,
working with pairs of fermion or boson operators without bosonization,
allows to better respect the Pauli principle and in the case of fermions
this has given excellent results in a series of applications to
non-trivial models where comparison with exact solutions was available
\cite{DS90,curu,duke}. There is no reason not to believe that this advantage
should not carry over to the case of relativistic interacting boson fields. In
the relativistic case, however, our ansatz \ref{DEA29} for the Green's function
has the disadvantage that it is not manifestly covariant and this is
certainly at the origin of some difficulties  we encountered in this
respect and which we have discussed in the main text. However, as we
have shown in section 4, even standard RPA has, in our model at least,
some difficulties with co-variance. Further studies are necessary to
elucidate this point and eventually to cure it. We believe nevertheless
that the results we have found in the present study are encouraging and
that we will be able to apply our theory in future work as successfully
to relativistic field theory as we did already in the past  for the
non-relativistic many body problem.

%%%%%%%%%%%%%%%%%%%%%%%%%%%%%%%%%%%%%%%%%%%%%%%%%%%%%%%%%%%%%%%%%%%%%%%%%%%%%%%
\section{Conclusion}
%%%%%%%%%%%%%%%%%%%%%%%%%%%%%%%%%%%%%%%%%%%%%%%%%%%%%%%%%%%%%%%%%%%%%%%%%%%%%%%

In this work we have tried to make the first step in elaborating an extension of RPA
theory, which has been very successful in the context of non-relativistic many body
problem \cite{R68,H64,CDS94,TS95,duke}, to relativistic field theory. 
Applications of standard RPA to relativistic
field theory has emerged in the recent past and proven its great potential interest
\cite{DMS96,ACSW96,ASW97}. The main quality of the RPA approach is to sum a certain
class of diagrams (the rings) to all order and by the same token to restore
spontaneously broken symmetries and to fulfill the conservation laws. The drawback of
standard RPA is to ignore, at a certain step of its derivation, correlations in the
vacuum (the quasi-boson approximation). This often entails a rather strong
overbinding of the ground state. To avoid this approximation is the aim of the afore
mentioned extension of RPA leading to the so-called Self-Consistent RPA (SCRPA). An
intermediate but considerably less complicated version of this theory is the so-called
renormalized RPA (r-RPA) where, with respect to standard RPA, only the occupation
numbers are modified due to ground state correlations. It is this latter version which
we have tried to develop here in the context of relativistic field theory with
application to the $\varphi^4$ theory in $1+1$ dimension. We have studied the transition to
a symmetry broken phase in varying the coupling constant. We have found a very slight
first order phase transition and concluded that it will turn to second order, as it is
expected in this model, once further correlations of the SCRPA are included.
This opinion stems from the fact that going from the mean field theory (Gaussian
approximation) to the r-RPA solution the first order character of the transition has
been very much attenuated. We also point out a certain number of difficulties with the
extension of RPA to relativistic field theory for bosons. 
This concerns for instance the fact that
the approach is not manifestly covariant. Although the standard RPA yields at
the end a covariant solution for the single particle mode, surprisingly we found that it has
difficulties for the calculation of the correlation energy with respect to covariance. 
Apparently this had not been noticed before. 
At the r-RPA level we find that covariance is
violated already for the single-particle mode and we have to restore it by an ad-hoc but natural 
assumption. It will be an interesting
further study whether SCRPA inherently violates covariance or whether this  is due
to the approximations we have been forced to introduce.   Another open question to be
studied in the future concerns renormalization. In the present model study 
this difficulty was absent since the $\varphi^4$ theory in $1+1$ dimensions is super
renormalizable. However, in the general case, this problem has obviously to be
mastered. Finally a detailed comparison of the diagrammatic content of the RPA and the cluster
expansion \cite{HCPT95}, in the context of relativistic bosonic
 theories with a broken symmetry,  would be certainly of great interest.

\noindent
In short we have applied for the first time an extension of RPA theory, which turns out
to be successful in the non relativistic many body problem, to a relativistic but
schematic field theoretic model. Although some problems are still present, 
we believe that our results are quite encouraging. Studies for the resolution of 
the remaining problems are under way.  

\section{Acknowledgments}
 We would like to thank Z.Aouissat for very useful discussions
concerning various aspects of  SCRPA and its application to field theoretical
problems.

%%%%%%%%%%%%%%%%%%%%%%%%%%%%%%%%%%%%%%%%%%%%%%%%%%%%%%%%%%%%%%%%%%%%%%%%%%%%%%%%%%%%
\section{Appendix}
%%%%%%%%%%%%%%%%%%%%%%%%%%%%%%%%%%%%%%%%%%%%%%%%%%%%%%%%%%%%%%%%%%%%%%%%%%%%%%%%%%

In this appendix we list the explicit expressions for the  Green's functions. 
For this purpose we introduce various quantities~:
\begin{eqnarray}
I^{(1)}_{\beta\beta'}(E)& = &{ 1\over 2\varepsilon_\beta} {1\over 2\varepsilon_{\beta'}}
\,{\varepsilon_\beta\,{\cal N}_\beta\,+\,\varepsilon_{\beta'}\,{\cal N}_{\beta'}\over
E\,-\,\varepsilon_\beta \,-\,\varepsilon_{\beta'}\,+\,i\eta}\nonumber\\
I^{(2)}_{\beta\beta'}(E)& = &{\varepsilon_\beta\,-\,\varepsilon_{\beta'} 
\over 2\,\varepsilon_\beta \,\varepsilon_{\beta'}}\,
{\varepsilon_\beta\,{\cal N}_\beta\,-\,\varepsilon_{\beta'}\,{\cal N}_{\beta'}\over
E^2\,-\,(\varepsilon_\beta \,-\,\varepsilon_{\beta'})^2\,+\,i\eta}\nonumber\\
I^{(3)}_{\beta\beta'}(E)& = &-{ 1\over 2\varepsilon_\beta} {1\over 2\varepsilon_{\beta'}}
\,{\varepsilon_\beta\,{\cal N}_\beta\,+\,\varepsilon_\beta'\,{\cal N}_{\beta'}\over
E\,+\,\varepsilon_\beta \,+\,\varepsilon_{\beta'}\,-\,i\eta}~.\nonumber\\
\end{eqnarray}
We also introduce the loop integrals~;
\begin{eqnarray}
I^{(i)}_\alpha (E) &=& {1\over V}\sum_{\beta{\beta'}}\,\delta_{\alpha-\beta-\beta'}\,
I^{(i)}_{\beta\beta'}(E)\nonumber\\
I_\alpha (E) &=& I^{(1)}_\alpha (E)\,+\,I^{(2)}_\alpha (E)\,+\,I^{(3)}_\alpha (E)~.
\end{eqnarray}
In particular for $\alpha$ corresponding to the momentum ${\vec P}$ one has the explicit
expression:
\begin{eqnarray}
I(E,{\vec P}) &\equiv & I_{\alpha={\vec P}}(E)\nonumber\\
& = &\int \,{d{\vec k}_1\,d{\vec k}_2\over (2\pi)^d}\,\delta^{(d)}\left({\vec P}-
{\vec k}_1 - {\vec k}_2\right)\,\bigg[
{\varepsilon_1\,+\,\varepsilon_2\over 2\,\varepsilon_1\,\varepsilon_2}\,
{\varepsilon_1\,{\cal N}_1\,+\,\varepsilon_2\,{\cal N}_2\over 
E^2\,-\,\left(\varepsilon_1\,+\,\varepsilon_2\right)^2+i \eta}\nonumber\\
& &-\,
{\varepsilon_1\,-\,\varepsilon_2\over 2\,\varepsilon_1\,\varepsilon_2}\,
{\varepsilon_1\,{\cal N}_1\,-\,\varepsilon_2\,{\cal N}_2\over 
E^2\,-\,\left(\varepsilon_1\,-\,\varepsilon_2\right)^2+i \eta}\bigg].
\end{eqnarray}
For the one-particle Green's functions one obtains~:
\begin {eqnarray}
G_{\alpha\alpha'^\dagger}(E) &=&\delta_{\alpha,\alpha'}\, { E\,+\varepsilon_\alpha\,+\,
{\Sigma_\alpha(E)\over 2\varepsilon_\alpha}\over 2\varepsilon_\alpha}
\,G_\alpha(E)\nonumber\\
G_{-\alpha^\dagger -\alpha'}(E) &=&\delta_{\alpha,\alpha'}\, { -E\,+\varepsilon_\alpha\,+\,
{\Sigma_\alpha(E)\over 2\varepsilon_\alpha}\over 2\varepsilon_\alpha}
\,G_\alpha(E)\nonumber\\
G_{-\alpha^\dagger \alpha'^\dagger}(E) &=& G_{\alpha -\alpha'}(E)= 
\delta_{\alpha,\alpha'}\, 
{\Sigma_\alpha(E)\over 4\varepsilon^2_\alpha}\, \,G_\alpha(E) ,
\end{eqnarray}
where the full propagator is~: 
\begin{equation}
G_{\phi_\alpha \phi^\dagger_{\alpha'}}(E)=\delta_{\alpha,\alpha'}\,G_\alpha(E)=
\delta_{\alpha,\alpha'}\,\left(E^2\,-\varepsilon^2_\alpha\,-\Sigma_\alpha(E)\right)^{-1}.
\end{equation}
The mass operator being given by~:
\begin{equation}
\Sigma_\alpha(E)={b^2\,s^2\over 2}{I_\alpha(E)\over 
1-{b\over 2}\,I_\alpha(E)}~.
\end{equation}

\smallskip\noindent
For what concerns the 2p-1h and 2p-2p Green's functions we introduces indices $i$ to label 
the destruction (creation) operators: $1=\beta, \beta' (\beta^\dagger, \beta'^\dagger)$, 
$2=(\beta, -\beta'^\dagger)_{sym} ((\beta^\dagger, -\beta')_{sym})$ and  
$3=-\beta^\dagger,  -\beta'^\dagger (-\beta, -\beta')$.
The results are~: 
\begin{eqnarray}
G^{(i)}_{\beta\beta', \alpha^\dagger}(E) &=& 
G^{(i)}_{\alpha^\dagger ,\beta\beta'}(E)\nonumber\\
&=& {b\,s\over \sqrt V}\,\delta_{\alpha-\beta-\beta'}\,
{I^{(i)}_{\beta\beta'}(E)\over 1-{b\over 2}\,I_\alpha(E)}\,
\left({E\,+\,\varepsilon_\alpha\over 2\,\varepsilon_\alpha}\right)\,
G_{\phi_\alpha \phi^\dagger_\alpha}(E)\nonumber\\
G^{(i)}_{\beta\beta', -\alpha}(E) &=& 
G^{(i)}_{-\alpha ,\beta\beta'}(E)\nonumber\\
&=& {b\,s\over \sqrt V}\,\delta_{\alpha-\beta-\beta'}\,
{I^{(i)}_{\beta\beta'}(E)\over 1-{b\over 2}\,I_\alpha(E)}\,
\left({-E\,+\,\varepsilon_\alpha\over 2\,\varepsilon_\alpha}\right)\,
G_{\phi_\alpha \phi^\dagger_\alpha}(E)
\end{eqnarray}

\begin{eqnarray}
G^{(i j)}_{\beta\beta', \,\gamma\gamma'}(E) &=& 
I^{(i)}_{\beta\beta'}(E)\,\delta_{i,j}\,\left(\delta_{\beta\gamma}\,
\delta_{\beta'\gamma'}\,+\,\delta_{\beta\gamma'}\,\delta_{\beta'\gamma}\right)\nonumber\\
& &+\,{b\over V}\,\sum_{\alpha}\,
{\delta_{\alpha-\beta-\beta'}I^{(i)}_{\beta\beta'}(E)\quad
\delta_{\alpha-\gamma-\gamma'}I^{(i)}_{\gamma\gamma'}(E)\over 
1-{b\over 2}\,I_{\alpha}(E)}\nonumber\\
& &+\,{b^2\, s^2\over V}\,\sum_{\alpha}\,
{\delta_{\alpha-\beta-\beta'}I^{(i)}_{\beta\beta'}(E)\quad
\delta_{\alpha-\gamma-\gamma'}I^{(i)}_{\gamma\gamma'}(E)\over 
\left(1-{b\over 2}\,I_{\alpha}(E)\right)^2}\, 
G_{\phi_{\alpha} \phi^\dagger_{\alpha}}(E).
\end{eqnarray}


\begin{thebibliography}{99}

\bibitem{RS80} P. Ring and P. Schuck, The Nuclear Many-Body Problem, Springer-Verlag
(1980).

\bibitem{S84} P.M. Stevenson,  Phys. Rev. D30 (1984) 1712
P.M. Stevenson,  Phys. Rev. D32 (1985) 1389,
 Y. Birhaye, M. Consoli, Phys. Lett. B157 (1985) 48,
P.M. Stevenson, Z. Phys. C35 (1987) 467,
P. M. Stevenson, B. All\`{e}s, R. Tarrach, Phys. Rev. D35 (1987) 2407,
A. K. Kerman, C. Martin and D. Vautherin, Phys.Rev. D47 (1993) 632, 
 G. Amelino-Camelia, Phys. Lett. B407 (1997) 268,
and references therein.

\bibitem{KV89} A. K. Kerman, D, Vautherin, Ann. Phys. (N.Y.) 192 (1989) 408
C. Martin, D. Vautherin, hep-ph/9401261,
I. I. Kogan, A. Kovner, Phys. Rev. D52 (1995) 3719, and references
therein.

\bibitem{DMS96} V. Dmitrasinovic, J. A. McNeil, J. R. Shepard,
Z. Phys. C 69 (1996) 35.

\bibitem{ACSW96} Z. Aouissat, G. Chanfray, P. Schuck and J. Wambach, Nucl. Phys. A603
(1996) 458. 
       
\bibitem{ASW97} Z. Aouissat, P. Schuck and J. Wambach, Nucl. Phys. A618 (1997) 402.

\bibitem{HCPT95} J.M. H\"{a}auser, W. Cassing, A. Peter and M.H. Thoma, Z. Phys. A353, 
(1995) 301. 

\bibitem{JS94} W. Janke and T. Sauer, Nucl. Phys. Proc. Suppl. 42 (1995) 917.

\bibitem{LW97} W. Loinaz and R.S. Willey, Phys.Rev. D58 (1998) 076003.     

\bibitem{MRL99} P.J. Marrero, E.A. Roura and D. Lee, Phys. Lett. B471 (1999) 45.

\bibitem{FW} A.L. Fetter and J.D. Walecka, Quantum theory of Many-particle Systems,
McGraw-Hill, New-York (1971). 

\bibitem{KER} A.K. Kerman,   Ann. of Phys,  269 (1998) 55.
      
\bibitem{R68} D.J. Rowe, Rev. Mod. Phys. 40 (1968) 153 and Nuclear Collective Motion, 
Methuen, London (1970).

\bibitem{DS90} J. Dukelsky and P. Schuck, Nucl. Phys. A512 (1990) 466.

\bibitem{H64} K. Hara, Progr. Theor. Phys. 32 (1964) 88.

\bibitem{CDS94} F. Catara, N. Dinh Dang and M. Sambatoro, Nucl. Phys. A579 (1994) 1.

\bibitem{TS95} J. Toivanen and J. Suhonen, Phys. Rev. Lett 75 (1995) 410.

\bibitem{SE73} P. Schuck and S. Ethofer, Nucl. Phys. A212 (1973) 269. 

\bibitem{curu} P. Curutchet, J. Dukelsky, G. G. Dussel, A. J. Fendrik, Phys.
Rev. C40 (1989) 2361.

\bibitem{bijk} R. Bijker, S. Pittel, J. Dukelsky, Phys. Lett. B219 (1989) 5.

\bibitem{blai} J. P. Blaizot, G. Ripka, {\it Quantum theory of finite systems},
(Mit Press, 1986)

\bibitem{ripk} G. Ripka, {\it Quarks bound by chiral fields. The quark structure
of the vacuum and of light mesons and baryons}, (Clarendon Press, 1997)

\bibitem{duke} J. Dukelsky, P.Schuck, Mod. Phys. Lett. A26 (1991) 2429; \\
P. Kr\"{u}ger, P. Schuck, Europhys. Lett. 72 (1994) 395; \\
J. Dukelsky, P.Schuck, Phys. Lett. B387 (1996) 233; \\
J. Dukelsky, G. R\"{o}pke, P.Schuck, Nucl. Phys. A 628 (1998) 17; \\
J. G. Hirsch, A. Mariano, J. Dukelsky, P.Schuck, nucl-th/0109036.

\bibitem{CDOS02} G. Chanfray, D. Davesne, M. Oertel and P. Schuck, work in progress.

\end{thebibliography}
\end{document}